\documentclass[reprint, superscriptaddress]{revtex4-1}
\usepackage{amsmath, amssymb, amsfonts}
\usepackage{bm}
\usepackage{color}
\usepackage{float}
\usepackage{graphicx}
\usepackage{subfig} 
\usepackage{dcolumn}
\usepackage{fixltx2e}
\usepackage{gensymb} 
\usepackage{textgreek} 
\usepackage{natbib} 

\graphicspath{{images/}{images_old/} }
\usepackage{color}

\begin{document}

\title{Influence of Neighborhood SES on Functional Brain Network Development}

\author{Ursula A. Tooley} 
\affiliation{Department of Neuroscience, Perelman School of Medicine, University of Pennsylvania, Pennsylvania, PA 19104, USA}
\author{Allyson P. Mackey} 
\affiliation{Department of Psychology, College of Arts and Sciences, University of Pennsylvania, Pennsylvania, PA 19104, USA}
\author{Rastko Ciric} 
\affiliation{Department of Psychiatry, Perelman School of Medicine, University of Pennsylvania, Pennsylvania, PA 19104, USA}
\author{Kosha Ruparel} 
\affiliation{Department of Psychiatry, Perelman School of Medicine, University of Pennsylvania, Pennsylvania, PA 19104, USA}
\author{Tyler M. Moore} 
\affiliation{Department of Psychiatry, Perelman School of Medicine, University of Pennsylvania, Pennsylvania, PA 19104, USA}
\author{Ruben C. Gur} 
\affiliation{Department of Psychiatry, Perelman School of Medicine, University of Pennsylvania, Pennsylvania, PA 19104, USA}
\author{Raquel E. Gur} 
\affiliation{Department of Psychiatry, Perelman School of Medicine, University of Pennsylvania, Pennsylvania, PA 19104, USA}
\author{Theodore D. Satterthwaite}
\affiliation{Department of Psychiatry, Perelman School of Medicine, University of Pennsylvania, Pennsylvania, PA 19104, USA}
\author{Danielle S. Bassett}
\affiliation{Department of Bioengineering, School of Engineering and Applied Sciences, University of Pennsylvania, Pennsylvania, PA 19104, USA}
\affiliation{Department of Neurology, Perelman School of Medicine, University of Pennsylvania, Pennsylvania, PA 19104, USA}
\affiliation{Department of Physics \& Astronomy, College of Arts and Sciences, University of Pennsylvania, Pennsylvania, PA 19104, USA}
\affiliation{Department of Electrical \& Systems Engineering, School of Engineering and Applied Sciences, University of Pennsylvania, Pennsylvania, PA 19104, USA}

\email{dsb@seas.upenn.edu}

\date{\today}

\begin{abstract}
Higher socioeconomic status (SES) in childhood is associated with increased cognitive abilities, higher academic achievement, and decreased incidence of mental illness later in development. Accumulating evidence suggests that these effects may be due to changes in brain development induced by environmental factors. While prior work has mapped the associations between neighborhood SES and brain structure, little is known about the relationship between SES and intrinsic neural dynamics. Here, we capitalize upon a large community-based sample (Philadelphia Neurodevelopmental Cohort, ages 8--22 years, $n=1012$) to examine developmental changes in functional brain network topology as estimated from resting state functional magnetic resonance imaging data. We quantitatively characterize this topology using a local measure of network segregation known as the clustering coefficient, and find that it accounts for a greater degree of SES-associated variance than meso-scale segregation captured by modularity. While whole-brain clustering increased with age, high-SES youth displayed faster increases in clustering than low-SES youth, and this effect was most pronounced for regions in the limbic, somatomotor, and ventral attention systems. The effect of SES on developmental increases in clustering was strongest for connections of intermediate physical length, consistent with faster decreases in local connectivity in these regions in low-SES youth, and tracked changes in BOLD signal complexity in the form of regional homogeneity. Our findings suggest that neighborhood SES may fundamentally alter intrinsic patterns of inter-regional interactions in the human brain in a manner that is consistent with greater segregation of information processing in late childhood and adolescence.
\end{abstract}

\maketitle

\section*{Introduction}

	Higher SES during youth is associated with lower risk for psychiatric disorders \cite{duncan_economic_1994, evans_childhood_2014}, increased executive function \cite{noble_socioeconomic_2007}, higher levels of educational attainment and income \cite{ryan_childhood_2006, mcloyd_socioeconomic_1998, duncan_importance_2012}, and better physical health \cite{cohen_childhood_2010, evans_childhood_2016}. SES is also associated with cortical development as early as infancy \cite{hanson_family_2013, tomalski_socioeconomic_2013, brito_associations_2016, betancourt_effect_2016, farah_neuroscience_2017, jha_environmental_2018}. Notably, emerging evidence points to a pattern of accelerated structural brain development in low-SES individuals \cite{piccolo_age-related_2016, lewinn_sample_2017}, suggesting that SES may influence not only the initial or final states of brain development \cite{jednorog_influence_2012}, but also the trajectory between them. Yet whether such differential trajectories also exist in functional brain development remains unknown. Findings of more protracted structural brain development are at odds with the overall impression of findings from studies of brain function, which have been performed at specific ages but not yet as a continuous function of development. These single-shot studies, on the whole, suggest potentially faster functional brain development in high-SES children and adults, including increased functional specialization in language regions in high-SES kindergartners \cite{raizada_socioeconomic_2008}, more mature frontal gamma power in high-SES infants \cite{tomalski_socioeconomic_2013}, and increased resting-state functional connectivity in high-SES children and adults \cite{barch_effect_2016, marshall_socioeconomic_2018, sripada_childhood_2014, smith_positive-negative_2015} (although whether this is indicative of greater maturation is unclear). This apparent paradox motivates a thorough investigation into whether and how SES relates to functional brain network development in youth.
	
	Cortical networks become increasingly specialized and segregated with age \cite{stiles_basics_2010, satterthwaite_heterogeneous_2013, fair_functional_2009,grayson_development_2017}. Increasing evidence suggests that this age-dependent pattern of cortical segregation supports the normative maturation of cognitive function \cite{baum_modular_2017, wig_segregated_2017,gu_emergence_2015}. Yet, explicit studies of regional specialization and circuit segregation have traditionally been hampered by the dearth of computational techniques and methodological approaches that are appropriate for the study of spatially distributed interconnected systems. Recent advances in network neuroscience have met this need by drawing on mathematics, physics, and computer science to formalize a model of the brain as a network of interacting elements \cite{bassett_network_2017}. The network approach has recently been used to examine the influence of SES on functional brain networks in aging \cite{chan_socioeconomic_2018}. SES was found to moderate age-related declines in functional network segregation across the lifespan, such that lower SES individuals showed faster brain aging, i.e., earlier reductions in functional network segregation. Although links between SES and network segregation could reflect genetic differences rather than environmental influences, a recent twin study of adolescents showed that the heritability estimate for the clustering coefficient, a statistic which measures the potential for information segregation, is low, and therefore that it might be especially sensitive to environmental influences \cite{van_den_heuvel_genetic_2013}. Taken together, these studies suggest that environmental exposures associated with low SES might influence age-related changes in network dynamics, but nothing is yet known about how SES impacts network segregation during development.
	
\begin{figure*}	
	\includegraphics[width=.92\textwidth]{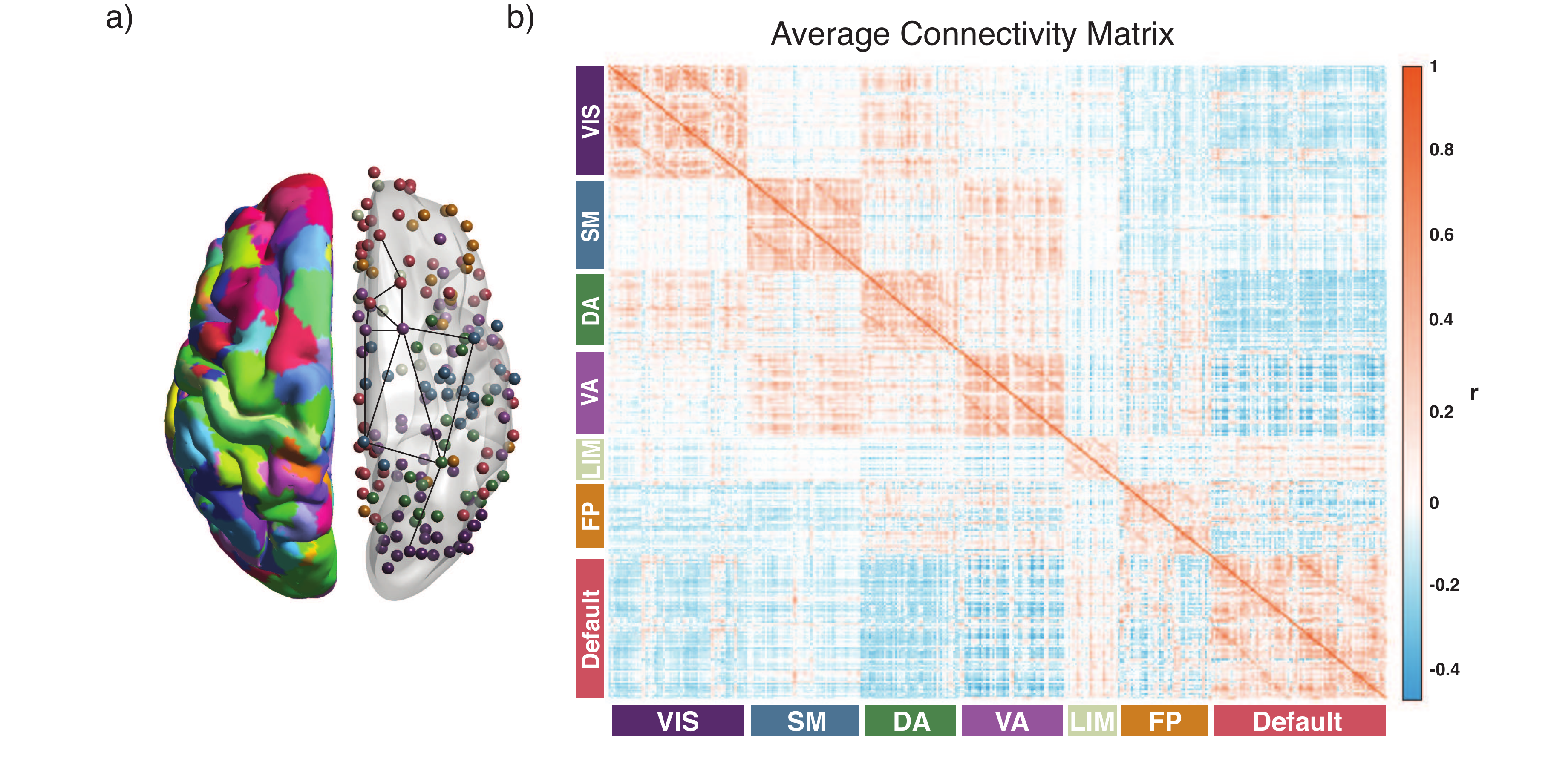}
\caption{\textbf{Schematic of approach.} \emph{(a)} A total of $N=360$ regions of interest in a multimodal cortical parcellation \cite{glasser_multi-modal_2016}. From each region we estimated the mean BOLD time series, and then we calculated the functional connectivity between any two regions using the Pearson correlation coefficient. For each subject, we collated all functional connectivity estimates into a single $N \times N$ adjacency matrix. The left hemisphere depicts the parcellation, while the right hemisphere depicts regions represented as network nodes and colored by their association to putative cognitive systems \cite{yeo_organization_2011}. \emph{(b)} The average functional connectivity matrix across subjects, ordered by putative cognitive systems: VIS: visual; SM: somatomotor; DA: dorsal attention; VA: ventral attention; LIM: limbic; FP: frontoparietal. }
	\label{fig0}
\end{figure*}

While previous studies lay important groundwork, a key gap remains, in that we have not yet probed the associations between neighborhood SES and trajectories of functional brain network segregation during childhood and adolescence. In the present report, we address this gap by leveraging the extensive cross-sectional neuroimaging data of the Philadelphia Neurodevelopmental Cohort (PNC), a community based sample of youth between the ages of 8 and 22 years \cite{satterthwaite_philadelphia_2016}. In light of prior work, we hypothesized that both the modularity quality index and the clustering coefficient would increase with age \cite{wu_topological_2013, satterthwaite_heterogeneous_2013, gu_emergence_2015} (although see Supekar et al. \cite{supekar_development_2009}). Moreover, based on recent evidence for decreased resting-state connectivity in low- as opposed to high-SES children and adults \cite{barch_effect_2016, sripada_childhood_2014}, we hypothesized that low-SES youth would exhibit either decreased segregation on average, or a slower increase in segregation with age, than high-SES youth. As described below, results provide novel evidence that neighborhood SES moderates age-related increases in segregation, as operationalized by a commonly studied metric in graph theory known as the clustering coefficient. Youth in high-SES neighborhoods had lower initial levels of local segregation and displayed larger increases in local segregation with age than youth in low-SES neighborhoods. This effect was partially explained by spatially distributed circuitry, indicated by middling-length connections. As a whole, our study provides the first evidence for the impact of SES on the development of functional network topology, providing greater insight into the neural manifestations of early environmental influences.

\section*{Results}

\begin{figure*}
	\includegraphics[width=0.9\textwidth]{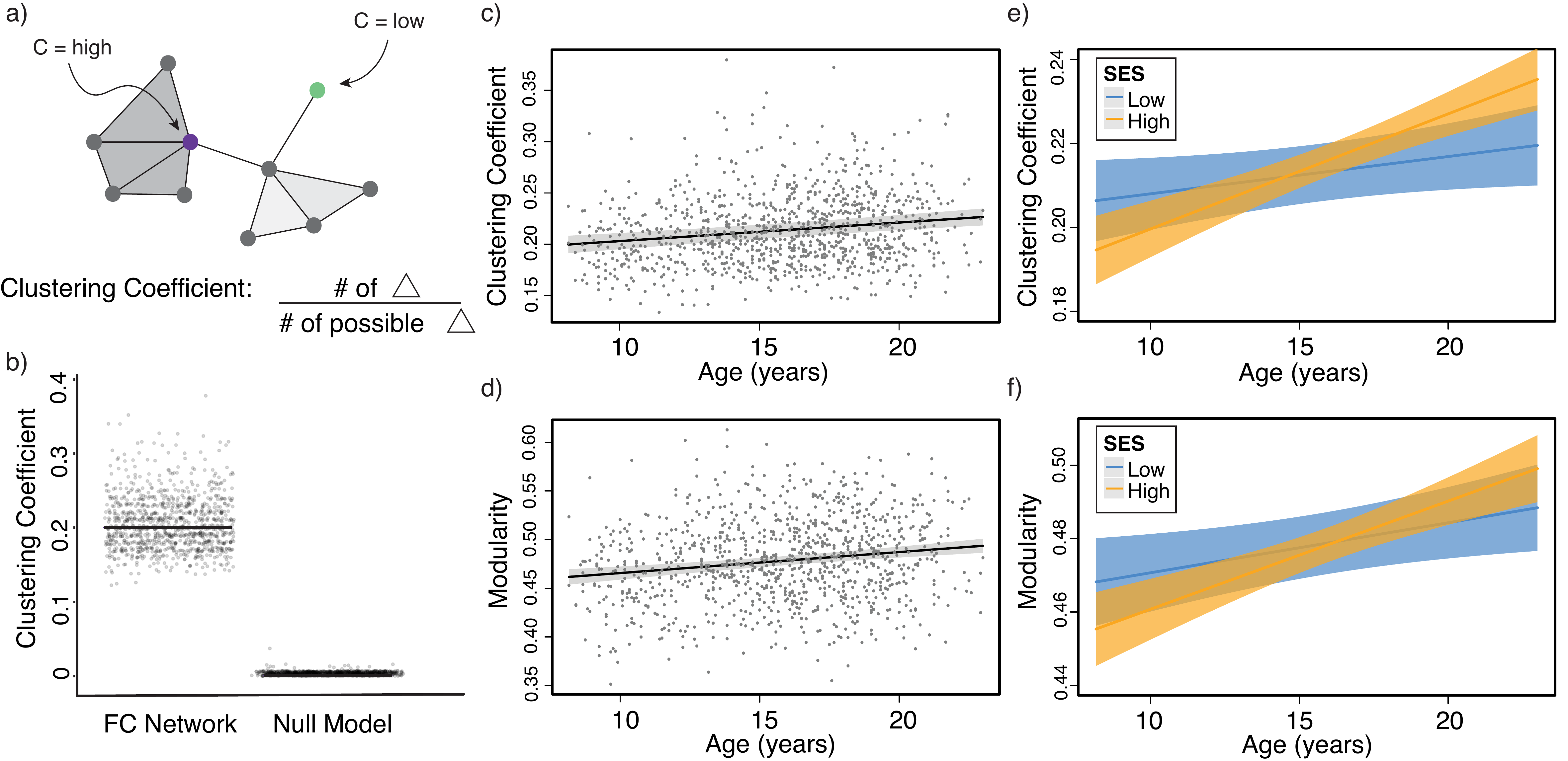}
\caption{\textbf{Effects of age and SES on whole-brain functional network topology at rest}. \emph{(a)} The clustering coefficient can be used to assess the degree to which neighboring nodes in a graph tend to cluster together. In a binary graph, such as that illustrated here, the clustering coefficient measures the fraction of triangles around a node; the purple node has a high clustering coefficient and the green node has a low clustering coefficient. In our study, we use an extension of this measure that is appropriate for signed, weighted networks. \emph{(b)} The average clustering coefficient is significantly higher in the observed functional brain networks than in random network null models in which the average edge weight, degree distribution, and strength distribution have been preserved ($p <1 \times 10^{-15}$). \emph{(c)} The average clustering coefficient increases with age, controlling for sex, race, head motion, and mean edge weight. Note that the values that are plotted are partial residuals. \emph{(d)} The value of the modularity quality index obtained by maximizing a modularity quality function (see Methods) also increases with age, controlling for sex, race, head motion, and mean edge weight. Note again that the values that are plotted are partial residuals. \emph{(e)} High-SES youth display a swifter increase in the average clustering coefficient of functional brain networks at rest than low-SES youth over this developmental period. \emph{(f)} High-SES youth also display a trend towards a swifter increase in the modularity quality index of functional brain networks at rest than low-SES youth over this developmental period. }
	\label{fig1}
\end{figure*}

In a total of $n=1012$ youth imaged as part of the PNC, we examine blood-oxygen-level-dependent (BOLD) signal in functional magnetic resonance images (fMRI) acquired as participants rested inside the scanner. We estimate inter-regional correlations in spontaneous, low-frequency BOLD time series, and we represent the inter-regional correlation matrix as a network, in which nodes represent brain regions and edges represent correlation values (\textbf{Fig.~\ref{fig0}b}). Next, we employ computational tools from network neuroscience to examine the effect of SES on functional brain network topology as a function of age. To parse topology, we focus on metrics of local-scale and meso-scale segregation that have been shown to change over development \cite{wu_topological_2013,satterthwaite_heterogeneous_2013, gu_emergence_2015, betzel_changes_2014, baum_modular_2017}: the clustering coefficient \cite{watts_collective_1998} (\textbf{Fig.~\ref{fig1}a}) and the modularity quality index \cite{newman_modularity_2006}, respectively. To characterize SES, we focus on neighborhood SES, reflecting the availability of social and community resources. Notably, though neighborhood SES is highly correlated with household-level SES, here we focus solely on the effect of neighborhood SES (above and beyond household-level SES, see Supplement S2). Neighborhood SES captures additional variance in the types of experiences children encounter, and has been shown to influence developmental outcomes above and beyond household-level SES \cite{chen_neighborhood_2014, chetty_effects_2016, marshall_socioeconomic_2018}. Neighborhood SES may become increasingly salient as children mature \cite{leventhal_neighborhoods_2000}; in this particular sample, neighborhood SES is more predictive of cognitive performance than parental characteristics such as education, race, or age \cite{moore_characterizing_2016}. 

	We take an explicitly hierarchical approach to the question of how functional brain network topology in development is affected by neighborhood SES. At the coarsest level of the hierarchy, we examine interactions between age and SES on whole-brain summary measures of network topology. At a median level of the hierarchy, we perform higher-resolution analysis of putative functional systems including the default mode, fronto-parietal, attention, and limbic systems, among others. At the lowest level of the hierarchy, we perform fine-grained analysis of individual brain regions. We then investigate several potential explanatory factors, including regional homogeneity (ReHo) of the BOLD signal as well as inter-regional Euclidean distance. To further confirm our results, we perform sensitivity analyses in a smaller sample ($n=883$) that excluded participants currently using psychoactive medication as well as participants with a history of psychiatric hospitalization.

\subsection*{Effects of age and SES on whole-brain functional network topology at rest}

We sought to understand how functional brain network topology at rest related to age and SES by first considering a metric of local segregation. We found that age significantly predicted the average clustering coefficient, with older children displaying higher average clustering coefficients than younger children (\textbf{Fig.~\ref{fig1}c}, $\beta = 0.17$, $p < 1 \times 10^{-6}$). Sex ($\beta = 0.10$, $p = 0.001$), mean edge weight ($\beta = 0.17$, $p < 1 \times 10^{-6}$), and in-scanner motion ($\beta = -0.16$, $p < 1 \times 10^{-6}$) were also significant predictors of average clustering coefficient. The main effect of SES was not significant ($\beta = 0.03$, $p =0.5$; full model: $F(7, 1004) = 16.07$, $R^{2} = 0.09$, $p < 1 \times 10^{-15}$). We observed no significant non-linear relationship between the average clustering coefficient and age (restricted likelihood ratio test $\mathrm{RLRT} = 0.39$ , $p > 0.15$) or network weight ($p > 0.2$).

Notably, we observed a significant interaction between SES and age, such that high-SES youth demonstrate a more rapid increase in average clustering coefficient with age than low-SES youth (\textbf{Fig.~\ref{fig1}e}, $p =0.004$). Mean edge weight ($\beta = 0.16$, $p < 1 \times 10^{-6}$) and motion ($\beta = -0.17$, $p < 1 \times 10^{-6}$) were also significant predictors of clustering coefficient in this model (full model: $F(8, 1003) = 15.18$, $R^{2} = 0.10$, $p < 1 \times 10^{-15}$). We observed similar results when controlling for maternal education, but do not find similar results when using maternal education in lieu of neighborhood SES (see Supplement S2). In a final conservative test, we observed no significant effects of age or age $\times$ SES interactions in random network null models that (i) preserved degree distribution, or (ii) preserved mean edge weight, degree distribution, and strength distribution (all $p$-values greater than 0.33). See Supplement S4 and \textbf{Sup. Fig. 4} for replication of these results in an alternative parcellation.

Next, using a metric of meso-scale segregation, we conducted a set of parallel analyses to determine the effect of age and SES on the modularity quality index. We observed that the modularity quality index increased with age (\textbf{Fig.~\ref{fig1}d}, $\beta = 0.16$, $p < 1 \times 10^{-6}$). Sex ($\beta = 0.10$, $p = 0.001$), mean edge weight ($\beta = -0.21$, $p < 1 \times 10^{-6}$), and in-scanner motion ($\beta = -0.07$, $p = 0.02$) were all significant predictors of the modularity quality index, the main effect of SES was not significant ($\beta = -0.01$, $p =0.8$; full model: $F(7, 1004) = 16.78$, $R^{2} = 0.10$, $p < 1 \times 10^{-15}$). Notably, a model including the interaction of age and SES was significant ($F(8, 1003) = 15.22$, $R^{2} = 0.10$, $p < 1 \times 10^{-15}$); the age-by-SES interaction was marginally significant (\textbf{Fig.~\ref{fig1}f}, $p = 0.05$). 

Intuitively, modules are composed of strong clusters. Thus, the trending significance of SES influences on developmental increases in modularity could possibly be driven by the strength of the effect on the clustering coefficient. Statistically, we observed that the modularity quality index is significantly correlated with the average clustering coefficient (Spearman's $\rho=0.78$, $p < 1 \times 10^{-15}$). To determine which effect was driving the observed results, we included both the average clustering coefficient and the modularity quality index in a single model. Even when thus controlling for modularity, we found that SES significantly moderated developmental increases in the average clustering coefficient (age $\times$ SES interaction, $p=0.03$). Interestingly, the reverse was not the case: when controlling for the average clustering coefficient, the moderating effect of SES on developmental increases in modularity was not significant (age $\times$ SES interaction, $p=0.5$).Thus, we cannot claim that changes in modularity are an important marker of the influence of SES on age-related changes in network topology, but rather conclude that the fundamental driver is the clustering coefficient and associated alterations in local network topology. 

\subsection*{Effects of age and SES on system-level functional network topology at rest}

\begin{figure}
	\includegraphics[width=0.5\textwidth]{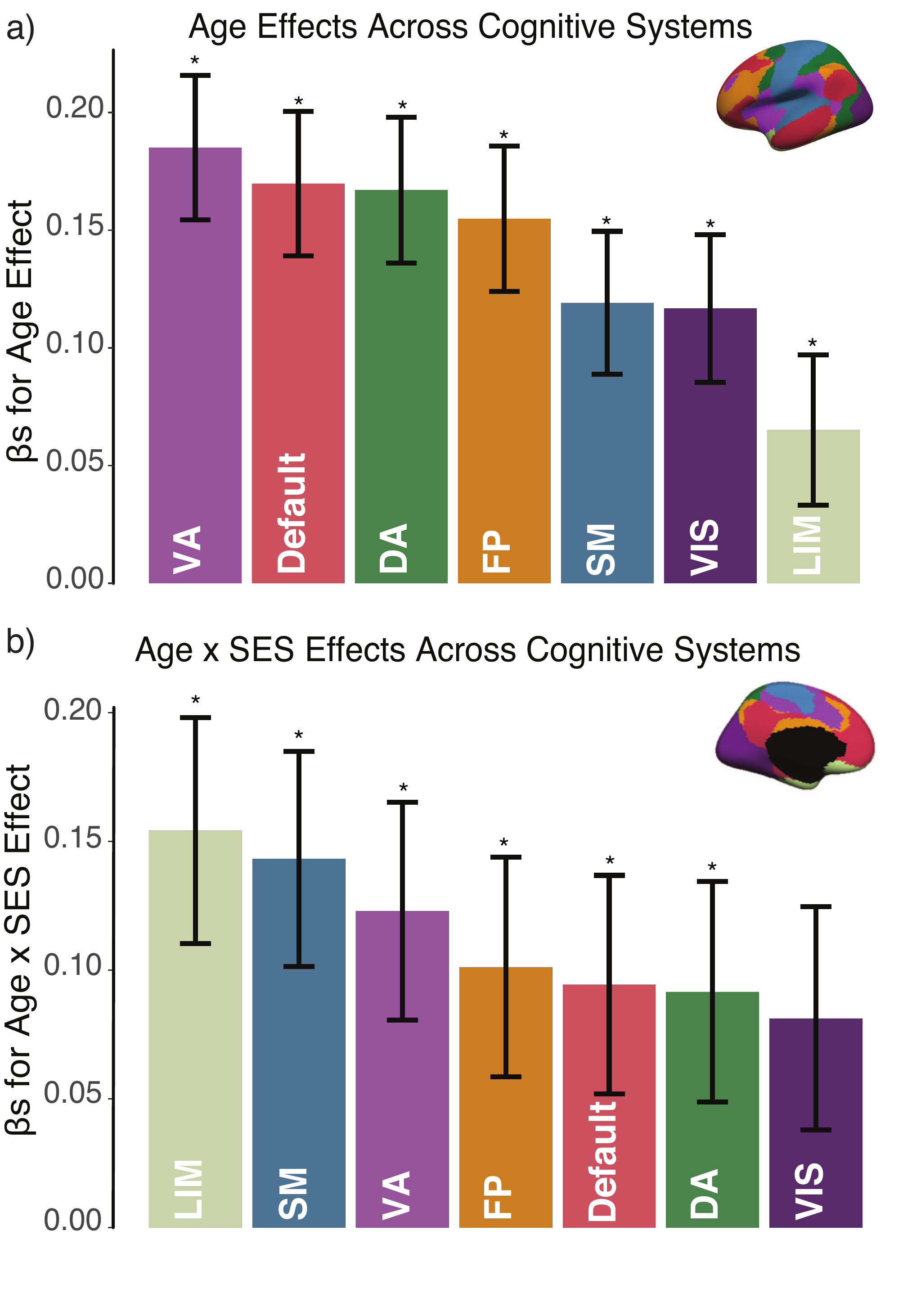}
	\caption{\textbf{Effects of age and SES on system-level functional network topology at rest.} \emph{(a)} Effects of age on increases in the clustering coefficient are strongest in the default mode, ventral attention, and dorsal attention systems. \emph{(b)} SES effects on age-related increases in the clustering coefficient were strongest in the limbic, somatomotor, and ventral attention systems. Error bars depict standard error of the mean, and asterisks indicate $p_{FDR} < 0.05$. VA: ventral attention; DA: dorsal attention. SM: Somatomotor. LIM: Limbic. FP: Frontoparietal. VIS: Visual.}
	\label{fig2}
\end{figure}

We next asked whether age-by-SES interactions might be particularly prevalent in one or more putative cognitive systems. To address this question, we assigned each brain region to one of seven systems defined \emph{a priori} \cite{yeo_organization_2011}, and we estimated the effect of age and SES on that system's regional clustering coefficients (\textbf{Fig.~\ref{fig2}}). We observed that developmental increases in the clustering coefficient varied across cognitive systems, with greatest increases in the default mode ($\beta = 0.17$, $p < 1 \times 10^{-7}$), ventral attention ($\beta = 0.19$, $p < 1 \times 10^{-8}$), and dorsal attention ($\beta = 0.17$, $p < 1 \times 10^{-7}$) systems. We also observed that the effect of SES on age-related increases in the clustering coefficient also varied, with the strongest age $\times$ SES interaction effects located in the somatomotor ($\beta = 0.14$, $p < 0.001$), limbic ($\beta = 0.15$, $p < 0.001$), and ventral attention ($\beta = 0.09$, $p = 0.03$) systems (see \textbf{Sup. Fig. 2} for full interaction plots by system). Finally, we calculated the $z$-score of the Rand coefficient between the vector of regional interaction $\beta$'s and the vector of regional system assignments, and found that the two were significantly more similar than would be expected by chance (permutation testing, $p < 0.0001$). Together, these data indicate that age-by-SES interactions are distinct across cognitive systems.

\subsection*{Effects of age and SES on region-level functional network topology at rest}

\begin{figure*}
	\includegraphics[width=.85\textwidth]{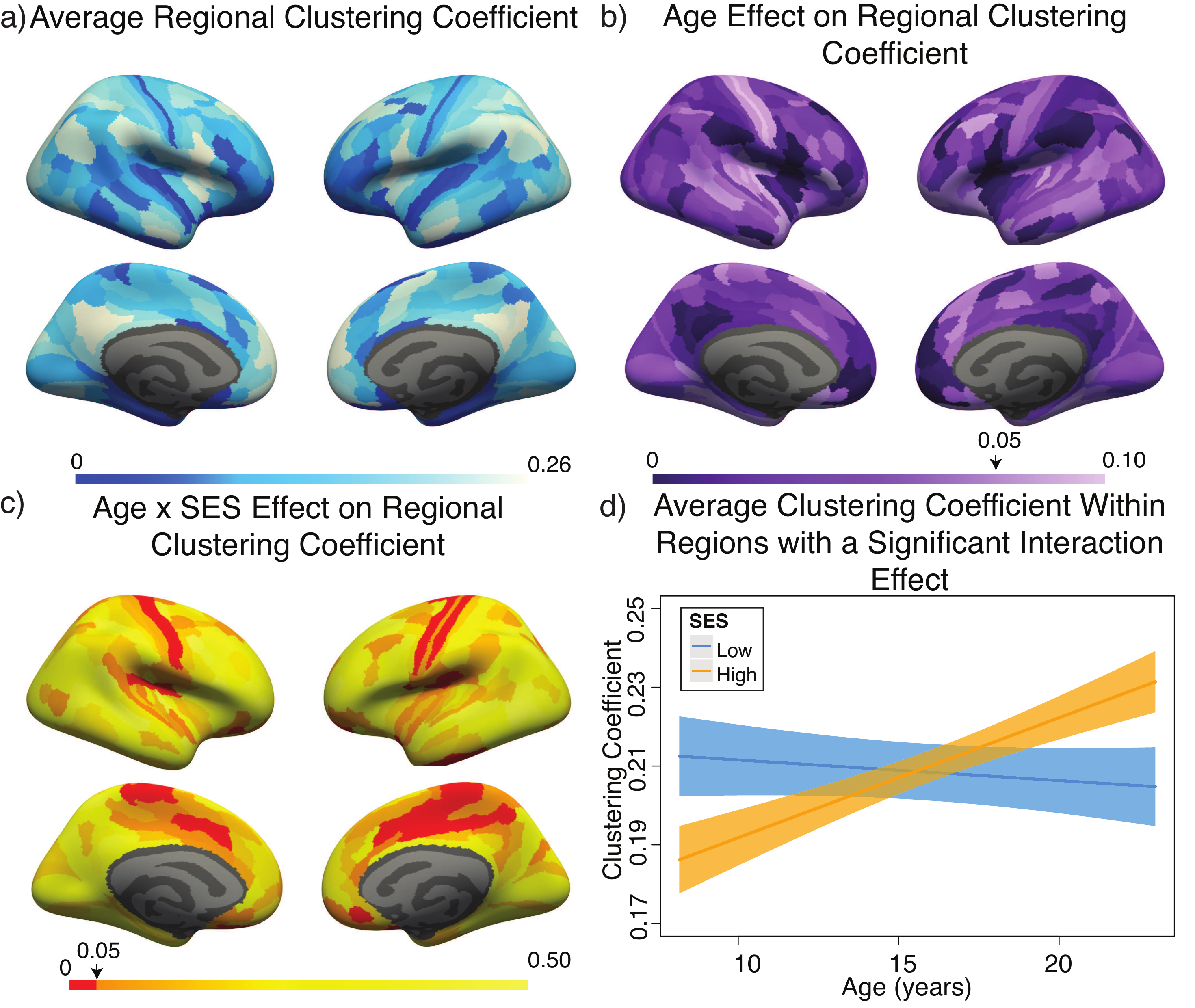}
	\caption{\textbf{Effects of age and SES on region-level functional network topology at rest.} \emph{(a)} On average, values of the clustering coefficient are largest in the precuneus, middle temporal gyrus, and inferior frontal gyrus. \emph{(b)} Regional $p$-values for the effect of age on the clustering coefficient. By visually comparing panel \emph{(b)} to panel \emph{(a)}, we observe that regions with high clustering coefficient tend to increase in clustering coefficient with age. \emph{(c)} Regional $p$-values for the effect of SES on age-related increases in the clustering coefficient. Significant age-by-SES interactions are located in the limbic, somatomotor, and ventral attention systems. Image is thresholded to control for multiple comparisons using an false discovery rate of $q< 0.05$; significant regions are shown in red. \emph{(d)} SES effects on age-related increases in the clustering coefficient, extracted from only nodes that show a significant age-by-SES interaction.}
	\label{fig3}
\end{figure*}

We next asked whether age-by-SES interactions might differ even within cognitive systems, at the level of single brain regions. To address this question, we first noted that the clustering coefficient was highest in the posterior cingulate, precuneus, middle temporal gyrus, and inferior frontal gyrus (\textbf{Fig.~\ref{fig3}a}). Age-related increases in the clustering coefficient were most salient in the orbitofrontal cortex, anterior cingulate, insular cortex, precuneus, and inferior parietal lobe (all $p_{FDR} < 0.05$; \textbf{Fig.~\ref{fig3}b}). The effects of SES on age-related increases in the clustering coefficient were strongest in anterior and posterior cingulate, orbitofrontal cortex, and somatomotor areas including precentral and postcentral gyri, and the paracentral lobule (all age $\times$ SES effects $p_{FDR} < 0.05$; \textbf{Fig.~\ref{fig3}c}). In these regions, high-SES youth display a swifter increase in the clustering coefficient with age than low-SES youth (\textbf{Fig.~\ref{fig3}d}).

\subsection*{Regional homogeneity}

To better understand these regional effects, we considered an even finer-grained measure of the neurophysiology: the regional homogeneity (ReHo), which assesses the similarity of the BOLD time course within a voxel to that of its neighboring voxels. We note that ReHo is calculated at the voxel level and averaged across voxels within each region separately, in contrast to the clustering coefficient which is calculated from edge weights reflecting functional connectivity between regions. Thus, ReHo and clustering coefficient are mathematically independent quantities, and any observed relation between them is not an artifact of the analysis, but rather could be indicative of underlying biological processes \cite{alexander-bloch_disrupted_2010, lee_linking_2017, zalesky_relationship_2012}. We observed that across participants, ReHo measurements averaged over the whole brain were not significantly correlated with the average clustering coefficient ($p = 0.17$, controlling for subject-level covariates of age, sex, race, and in-scanner motion). Notably, we found  that ReHo significantly decreased with age ($\beta = -0.14$, $p < 1 \times 10^{-5}$), but the main effect of SES was not significant ($p > 0.5$, see Supplement S5 for full models). In a model including the age-by-SES interaction, the interaction was not significant ($p > 0.33$).

\begin{figure*}
	\includegraphics[width=.9\textwidth]{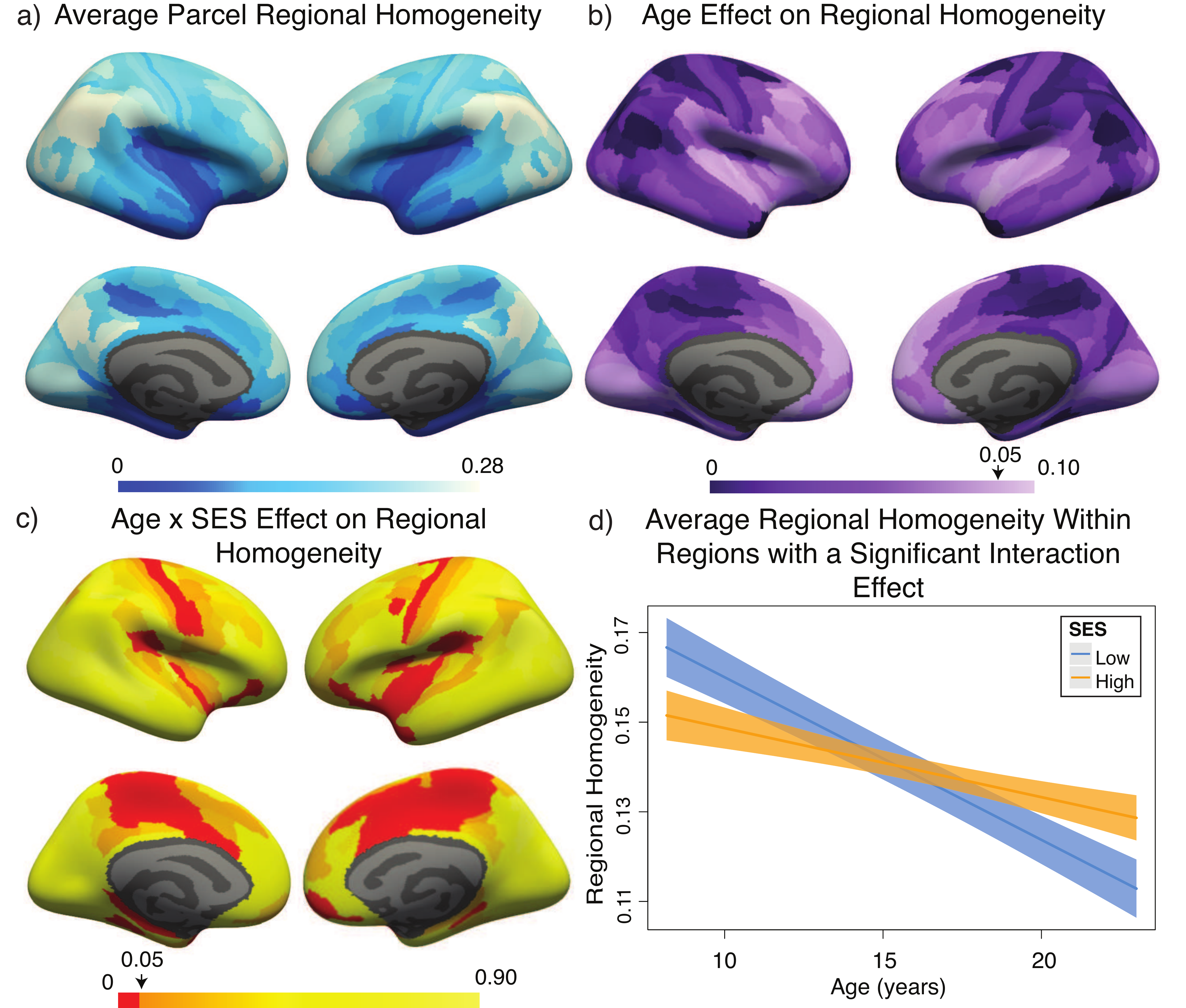}
	\caption{\textbf{Effects of age and SES on regional homogeneity of functional brain networks at rest.} \emph{(a)} On average, values of regional homogeneity are largest in the inferior parietal lobe, precuneus, and posterior cingulate. \emph{(b)} Regional $p$-values for the effect of age on regional homogeneity. Regional homogeneity decreases with age were widespread, and were strongest in the precuneus, inferior parietal lobe, and premotor cortex. \emph{(c)} Regional $p$-values for the effect of SES on age-related decreases in regional homogeneity. Image is thresholded to control for multiple comparisons using the false discovery rate ($q< 0.05$); significant regions are shown in red. \emph{(d)} SES effects on age-related decreases, extracted from only nodes that show a significant age-by-SES interaction.}
	\label{fig4}
\end{figure*}

In contrast to the whole-brain effects, we found that across regions, mean ReHo measurements were significantly correlated with the mean nodal clustering coefficient ($r=0.59$, $p < 1 \times 10^{-15}$), a relation that held even after controlling for subject-level covariates (all $p_{FDR}$'s$ < 0.00022$, \textbf{Sup. Fig. 5}). We examined region-level measurements of ReHo for age (\textbf{Fig.~\ref{fig4}b}) and age-by-SES effects (\textbf{Fig.~\ref{fig4}c}). We found that approximately a fifth of brain regions showed an SES effect on developmental decreases in ReHo (59 of 359 regions, age-by-SES interaction $p_{FDR} < 0.05$). In these regions, high-SES youth show initially lower ReHo, and ultimately later in development higher levels of ReHo than low-SES youth (\textbf{Fig.~\ref{fig4}d}, see Supplement S5 for full whole-brain models). Although the slope of the age effect on ReHo is reversed compared to that of the clustering coefficient (average ReHo $\beta = -0.23$), the directionality of the age-by-SES interaction is similar. 

It is interesting to note that while ReHo does not explain the whole-brain effect of SES on age-related increases in the clustering coefficient, it might at least partially explain the regional effects: low-SES youth display more swiftly decreasing regional homogeneity with age, which might lead to decreased edge weight due to reduced signal amplitude and thus weaker clustering of edges forming triangles connecting functionally related areas. By visually comparing \textbf{Fig.~\ref{fig3}c} to \textbf{Fig.~\ref{fig4}c}, we observe that regions with a significant age-by-SES interaction on the clustering coefficient are similar to those demonstrating an age-by-SES interaction on ReHo. Thus, we next regressed the regional effect of ReHo out of the clustering coefficient, and found that this procedure eliminates the age-by-SES interaction in predicting the clustering coefficient (all regional age $\times$ SES $p$-values greater than $0.05$). These results indeed suggest that ReHo partially explains regional effects of SES on developmental increases in the clustering coefficient.

\subsection*{Spatial embedding of the effects of age and SES on functional network topology at rest }

\begin{figure*}
	\includegraphics[width=0.65\textwidth]{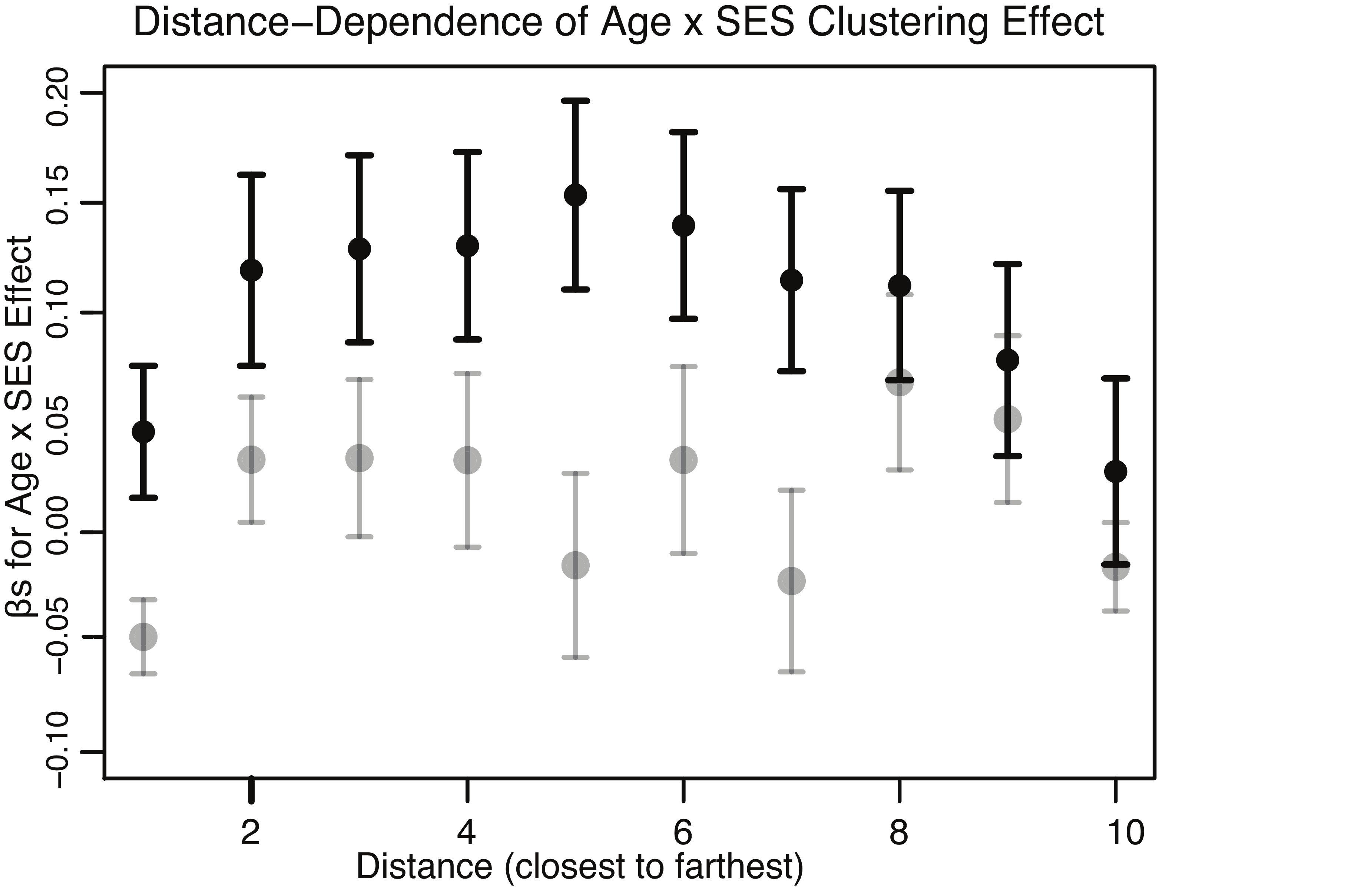}
	\caption{\textbf{Dependence of age-by-SES interactions on the length of functional connections.} We observed the strongest effects of SES on age-related increases in the clustering coefficient in subgraphs of the whole-brain functional network that were comprised of middling-length connections. Here we show $\beta$ values from the original model, calculated on subgraphs of the whole-brain functional network that are comprised of edges of different lengths. Error bars depict standard error of the mean. The $\beta$ values from models calculated on randomized null models of subgraphs are shown in gray. For simplicity, we segregated edges into 10 bins according to their lengths.} 
	\label{fig5}
\end{figure*}

The fact that the effect of SES on age-related increases in clustering coefficient are not fully explained by local homogeneity of connectivity as measured by ReHo, suggests the presence of a second explanation dependent on non-local, or spatially distributed processes. To probe this possibility, we separated edges into 10 equal-sized bins based on their physical length, as estimated by the Euclidean distance between regional centroids. For each distance bin, we constructed a subgraph of the full network that was only comprised of the edges whose physical distances were located within that bin, and we estimated the clustering coefficient on that subgraph. We observed the strongest effect of SES on age-related increases in the clustering coefficient in subgraphs composed of middle-length connections (\textbf{Fig.~\ref{fig5}}), consistent with the notion that these spatially distributed circuits are influenced by SES. The effects of SES on age-related increases in clustering coefficient were significantly greater in middling-length connections than expected in permutation-based null models ($p < 1 \times 10^{-15}$; see Methods). 

One possible corollary of this finding is that these middling-length edges could preferentially connect regions that show significant moderating effects of SES. To determine whether this was indeed the case, we estimated the age-by-SES effect on each edge separately, and conducted permutation tests on the distribution of age-by-SES effects across 3 groups of edges; (i) edges between nodes that demonstrated a significant node-level effect on the clustering coefficient, (ii) edges between these nodes and the rest of the brain, and (iii) edges connecting nodes that did not show a significant node-level effect on the clustering coefficient (see Methods). We observed that the effect of SES on age-related increases in edge strength was significantly stronger on edges between nodes that showed a significant age-by-SES effect on the regional clustering coefficient ($\beta =0.0460$) than on edges connecting those nodes to the rest of the network ($\beta =0.0055$, $p < 0.0001$) or on edges connecting nodes that did not show an age-by-SES effect ($\beta = 0.0012$, $p < 0.0001$). This observation supports the intuitive conclusion of our initial findings: the edges that are increasing the clustering coefficient more swiftly in high-SES youth than low-SES youth are in fact the medium-length edges forming triangles connecting regions that show a significant age-by-SES effect.

\subsection*{Sensitivity analysis}

We conducted sensitivity analyses with a smaller sample ($n=883$), excluding participants that were currently taking psychoactive medication or who had a history of psychiatric hospitalization. Results were qualitatively similar to those found in the full sample (see Supplement S3, \textbf{Sup. Fig. 3}). We observed a significant increase in the whole-brain clustering coefficient with age ($\beta = 0.17$, $p < 1 \times 10^{-6}$, full model:$F(7, 875) = 12.43$, $R^{2} = 0.08$, $p < 1 \times 10^{-14}$). We also observed a significant interaction between SES and age, such that higher-SES youth displayed a swifter increase in mean clustering coefficient with age than low-SES youth ($p = 0.004$; full model: $F(8, 874) = 12.01$, $R^{2} = 0.09$, $p < 1 \times 10^{-16}$).

\section*{Discussion}

Here we examined the effects of neighborhood SES on the development of functional brain network topology at rest. We took an explicitly multi-level approach, hierarchically investigating whole-brain summary measures of network topology, followed by higher-resolution analysis of functional systems and individual brain regions, as well as conducting analyses of metrics at several scales, from the meso-scale metric of the modularity quality index to the finer-grained metric of regional homogeneity. In a large community-based sample of 1012 youth, we uncovered evidence for an increase in local segregation of the network with age, as operationalized in a commonly studied metric in graph theory known as the clustering coefficient. Youth in high-SES neighborhoods had lower initial levels of local segregation and displayed larger increases in local segregation with age than youth in low-SES neighborhoods. Local segregation increased with age most strongly in the ventral attention, default mode, and dorsal attention systems, and SES influenced age-related increases in local segregation most in the limbic, somatomotor, and ventral attention systems. Importantly, the effects of SES on age-related increases in the clustering coefficient are partially explained by spatially distributed circuitry indicated by middling-length connections, and partially explained by changes in intra-regional connectivity in the form of regional homogeneity in the BOLD time course. Collectively, our results provide insight into the changes that occur in intrinsic functional brain network topology across development, and how the environment might bend these normative trajectories.

In youth ages 8--22 years, we found that local segregation as assessed by the clustering coefficient increased across development. This is, to our knowledge, the largest study to date to address changes in local segregation across this age range, and clarifies previous literature reporting both no change in local segregation with age \cite{supekar_development_2009, fair_functional_2009} and increases in local segregation with age \cite{wu_topological_2013} in functional brain networks. In our study, the observed increases in local segregation with age were highest in regions in the ventral attention and default mode systems, including areas of the frontal cortex, anterior cingulate, insular cortex, precuneus, and inferior parietal lobe. Increased network segregation with development has been widely reported in both structural and functional brain networks \cite{satterthwaite_heterogeneous_2013, fair_functional_2009, baum_modular_2017} and is believed to reflect the increasing segregation and refinement of modular network architecture with age \cite{grayson_development_2017}, a process that has been found to begin in infancy \cite{gao_development_2015} and continue through childhood and adolescence \cite{gu_emergence_2015}. There is increasing evidence that network segregation is advantageous for functional specialization, adaptability to task demands, and the reduction of interference across disparate functions \cite{fornito_competitive_2012, wig_segregated_2017}, and thus these neurodevelopmental changes in local network organization could contribute to supporting maturation of cognitive functions through childhood and adolescence \cite{baum_modular_2017}.

We found that neighborhood SES moderated age-related increases in local segregation, such that high-SES youth had faster increases in the clustering coefficient than low-SES youth, a pattern suggestive of faster functional brain development in the high-SES population. Previous work in a similar age range has suggested that high household SES is linked to a more protracted trajectory of structural brain development \cite{lewinn_sample_2017, piccolo_age-related_2016}. In light of our findings, this evidence suggests that SES might differentially affect functional and structural brain development, such that high-SES youth have a more protracted trajectory of structural brain development, but faster functional brain development.  Alternatively, our findings might not be at odds with findings of protracted structural development in high-SES children, as high-SES youth might initially show lower levels of local segregation in late childhood due to increased synaptogenesis and widespread connectivity early in development. This putative mechanism is also supported by findings of greater gray matter volume and faster growth trajectories in high-SES infants and children \cite{hanson_family_2013}, which would then rapidly develop into a more segregated network architecture as synapse elimination and pruning continues into adulthood \cite{huttenlocher_neural_2009, innocenti_exuberance_2005}. Our findings parallel a recent report that SES moderates age-related declines in functional network segregation, such that high-SES older adults show attenuated declines in segregation with age \cite{chan_socioeconomic_2018}. 

Another possibility is that neighborhood SES might have effects on brain development that are distinct from those of household SES, as we do not find a similar pattern of effects when using maternal education in lieu of neighborhood SES \cite{marshall_socioeconomic_2018, whittle_role_2017}. Our results add depth to previous findings of alterations in connectivity that report both increases \cite{marshall_socioeconomic_2018} and decreases \cite{barch_effect_2016, gao_functional_2015, sripada_childhood_2014} in resting state functional connectivity in low-SES populations. Our results are also in keeping with recent evidence of increased modularity in structural covariance networks of men from affluent neighborhoods as compared to deprived neighborhoods \cite{krishnadas_envirome_2013}; we also found that high-SES youth had faster increases in modularity with age than low-SES youth, although this effect was less strong than that on local segregation.

High functional local segregation indicates that a given node is connected with a group of densely interconnected local clusters \cite{bullmore_complex_2009}. Many advantageous topological properties of brain networks entail a tradeoff between these properties and wiring cost \cite{bullmore_economy_2012}. If the regions in which we see increases in functional local segregation were spatially confined, this increase would be parsimonious in the sense of physical wiring cost over shorter axonal distances \cite{henderson_geometric_2011}. In contrast, we observe that the effects of neighborhood SES on age-related increases in the clustering coefficient were strongest at mid-range connection lengths. Increases in clustering between more distant regions entails a higher biological cost \cite{heuvel_rich-club_2011}, and this might suggest that neighborhood SES entails a metabolic or environmental constraint on the more costly connections in the brain. The limbic system is particularly sensitive to experiences of adversity during development \cite{cameron_social_2017}, and in accordance with others' findings \cite{hanson_structural_2012, gianaros_parental_2011}, we find that prefrontal areas of the limbic system show the strong effects of neighborhood SES on increases in local segregation. (Note, however, that early adversity and neighborhood SES are overlapping but distinct constructs.) We also found strong effects of SES on increases in local clustering in the somatomotor network, especially in primary motor cortex, reminiscent of other evidence for the effects of SES on functional brain network development in somatomotor regions present as early as infancy \cite{gao_functional_2015}. These regions undergo earlier maturation \cite{deoni_investigating_2012, miller_prolonged_2012}, perhaps contributing to our ability to detect effects of environment in childhood and adolescence, and are situated at the far end of the sensorimotor-transmodal gradient of functional cortical organization \cite{margulies_situating_2016,huntenburg_large-scale_2018}.

We examined intra-regional local connectivity (regional homogeneity), an even finer measure of the local neurophysiology than local segregation, as a potential factor influencing our results \cite{jiang_regional_2016}. Zalesky et al. \cite{zalesky_relationship_2012} found that reductions in ReHo were associated with a reduction in the strength of edges connecting those regions, potentially due to reduced signal coherence and amplitude. In a similar vein, we found that local homogeneity of connectivity is correlated with local segregation across regions \cite{lee_linking_2017, alexander-bloch_disrupted_2010}. Moreover, we found that controlling for regional homogeneity accounts for the age-by-SES interaction effects on regional local segregation. Primary sensory networks including the somatomotor network have high inter-individual variability in regional homogeneity \cite{jiang_regional_2016}, which might contribute to the moderating effect of SES seen on decreases in regional homogeneity in somatomotor areas. Developmental changes in intra-regional homogeneity and network segregation have both been posited to be due to pruning of local connections \cite{jiang_toward_2015, lopez-larson_local_2011,lim_preferential_2015, supekar_development_2009, huang_development_2015}. However, we note that developmental trajectories of changes in intra-regional homogeneity are unlike those of local segregation, as in this age range the former decreases with age and the latter increases. Though not the focus of this paper, we speculate that this might represent a developmental shift from local connectivity to the segregated mid-range connections required for distributed cognitive operations \cite{fair_functional_2009, petersen_brain_2015, cohen_segregation_2016}.

\subsection*{Methodological considerations}

Several limitations inherent to this study are worth mentioning. First, this is a cross-sectional sample, with which we have limited power to examine developmental processes \cite{kraemer_how_2000}. Longitudinal studies examining environmental influences on the development of functional brain network topology will be necessary to further validate our results. Second, results with developmental samples have recently been shown to be disproportionately affected by head motion \cite{satterthwaite_impact_2012, roalf_impact_2016, byrge_identifying_2018}. To mitigate the impact of motion-related artifacts on our results, we applied current best practices in motion correction \cite{ciric_benchmarking_2017} and controlled for motion in all of our analyses. Third, we examined neighborhood-level SES as a composite factor tied to U.S. Census data by geocoding, an approach that aligns with the few studies that examine neighborhood SES and brain development \cite{krishnadas_socioeconomic_2013, whittle_role_2017}. However, SES is a multi-faceted construct with various components that might have differing associations with brain network organization, and thus a major goal for future research is to assess different indicators of SES and their independent contributions to effects on brain development \cite{farah_neuroscience_2017, ursache_neurocognitive_2016}. Fourth, we employed null network models to examine whether functional network topology differed significantly from that expected by chance in a random network of equivalent weight and degree distribution \cite{rubinov_weight-conserving_2011}, an approach with limitations when applied to correlation networks \cite{zalesky_use_2012}. The development of alternative null models that more closely approximate the architecture of correlation networks commonly found in biological data is an important area for future research \cite{barabasi_network_2004, bazzi_community_2016}.

\section*{Conclusions}

To our knowledge this study presents the first evidence of effects of neighborhood SES on resting state functional brain network topology in developing youth. We found that increases in local segregation with age are influenced by neighborhood SES, consistent with an interpretation of faster functional brain development in youth from high-SES neighborhoods. Neighborhood effects on brain development have been found above and beyond household-level effects \cite{whittle_role_2017, marshall_socioeconomic_2018}, and may be tied to inequalities in resource distribution or in perceived safety and cohesion of the neighborhood \cite{diez_roux_neighborhoods_2010}. Our findings add to the growing body of literature emphasizing the importance of the neighborhood environment during development, and suggests that neighborhood-level interventions for low-SES communities hold promise for promoting healthy brain development. 

\section*{Acknowledgments}
DSB and UAT acknowledge support from the John D. and Catherine T. MacArthur Foundation, the Alfred P. Sloan Foundation, the ISI Foundation, the Paul Allen Foundation, the Army Research Laboratory (W911NF-10-2-0022), the Army Research Office (Bassett-W911NF-14-1-0679, Grafton-W911NF-16-1-0474, DCIST- W911NF-17-2-0181), the Office of Naval Research, the National Institute of Mental Health (2-R01-DC-009209-11, R01-MH112847, R01-MH107235, R21-M MH-106799), the National Institute of Child Health and Human Development (1R01HD086888-01), National Institute of Neurological Disorders and Stroke (R01 NS099348), and the National Science Foundation (BCS-1441502, BCS-1430087, NSF PHY-1554488, BCS-1631550, and Graduate Research Fellowship to UAT). Additional support was provided by R01MH113550 (TDS \& DSB), R01MH107703 (TDS), R21MH106799 (DSB \& TDS), R01MH107235 (RCG), and the Penn/CHOP Lifespan Brain Institute. The content is solely the responsibility of the authors and does not necessarily represent the official views of any of the funding agencies. 

\clearpage
\newpage

\bibliography{rest_ses_networks_v1,rest_ses_networks_v1_2,rest_ses_networks_v1Notes}
\bibliographystyle{unsrt}

\newpage

\section*{Materials and Methods}

\subsection*{Contact for Reagent and Resource Sharing}
Further information and requests regarding resource sharing may be directed and will be fulfilled by the Lead Contact, Danielle S. Bassett (dsb@seas.upenn.edu).

\subsection*{Experimental Model and Subject Details}

\subsubsection*{Participant sample}

The Philadelphia Neurodevelopmental Cohort is a large community-based sample of youth between the ages of 8 years and 22 years, a subset of which participated in an extensive neuroimaging protocol ($n=1601$) \cite{calkins_philadelphia_2015}. After excluding youth with abnormalities in brain structure or a history of medical problems that could impact brain function ($n=154$), poor imaging data that did not pass quality assurance protocols (described in detail below, $n=432$), or inadequate coverage of some brain regions in our parcellation ($n=3$), we selected a subsample of $n=1012$ children and adolescents ($n_{f}=552$ female, mean age $15.78$ years) with neuroimaging data. Excluded participants were significantly younger than our analysis sample (mean age of included participants $15.78$, mean age of excluded participants $13.49$; student's $t$-test $p < 0.001$); socioeconomic status did not differ significantly between included and excluded participants ($p > 0.08$). Demographic information was collected by participant report, including sex, race (coded as White, Black, or Other; Other includes Asian, Native American, Hawaiian Pacific Islander, and Multi-Racial), and home address.

\subsubsection*{Measurement of socioeconomic status}

To represent participant neighborhood socioeconomic status, comprised of census-level data of the census block of each participant, we studied a factor score previously derived from this dataset \cite{moore_characterizing_2016}. This factor summarized the observed variance across multiple features including percent of residents married, percent of residents in poverty, median family income, percent of residents with a high school education, population density, and percent of residents employed. In the main text, we present several analyses with a median split of high and low neighborhood SES groups for ease of visualization and interpretation (see Supplement S1 for analyses with SES as a continous variable, results are qualitatively similar). In the high neighborhood SES group, mean percent in poverty was 5\%, mean percent married was 60\%, and mean median family income was \$107,039; in the low neighborhood SES group, mean percent in poverty was 27\%, mean percent married was 30\%, and mean median family income was \$41,946. The high and low neighborhood SES groups were not significantly different in age ($p > 0.22$) or sex ($p > 0.28$, see \textbf{Supplement Table II} for bivariate relationships between predictors).

\subsection*{Method Details} 

\subsubsection*{Imaging data acquisition}

All imaging data were acquired on the same 3T Siemens Tim Trio scanner with 32-channel head coil at the Hospital of the University of Pennsylvania.  Blood-oxygen-level-dependent (BOLD) signal was measured using a whole-brain, single-shot, multi-slice, gradient-echo (GE) echoplanar (EPI) sequence with the following parameters: 124 volumes; TR = 3000 ms; TE = 32 ms; flip angle=90 degrees, FoV = 192 $\times$ 192mm, matrix=64 $\times$ 64, slice thickness = 3 mm, slide gap = 0 mm, effective voxel resolution = 3 $\times$ 3 $\times$ 3 mm.  During the resting state scan, subjects were instructed to keep their eyes open and fixate on a white crosshair presented against a dark background.  Prior to the resting-state acquisition, a magnetization-prepared, rapid acquisition gradient-echo (MPRAGE) T1-weighted image was acquired to aid spatial normalization to standard atlas space, using the following parameters: TR = 1810 ms, TE = 3.51 ms, FOV = 180 $\times$ 240 mm, matrix = 256 $\times$ 192, 160 slices, TI = 1100 ms, flip angle = 9 degrees, and effective voxel resolution of 0.9 $\times$ 0.9 $\times$ 1 mm. Further study procedures and design are described in detail elsewhere \cite{satterthwaite_neuroimaging_2014}.

\subsubsection*{Imaging data preprocessing}

Whole-head T1 images were registered to a custom population template created with advanced normalization tools (ANTs) \cite{avants_reproducible_2011} using the top-performing diffeomorphic SyN registration \cite{klein_evaluation_2009}. Preprocessing of resting state time series was conducted using a validated confound regression procedure that has been optimized to reduce the influence of subject motion \cite{satterthwaite_improved_2013, ciric_benchmarking_2017}; preprocessing was implemented in XCP engine, a multi-modal toolkit that deploys processing instruments from frequently used software libraries, including FSL \cite{jenkinson_fsl_2012} and AFNI \cite{cox_afni:_1996}. Further documentation is available at \url{https://pipedocs.github.io/intro.html} and \url{https://github.com/PennBBL/xcpEngine}. Following distortion correction using a B0 map, the first 4 volumes of the functional timeseries were removed to allow signal stabilization, leaving 120 volumes for subsequent analysis. Functional timeseries were band-pass filtered to retain frequencies between 0.01 Hz and 0.08 Hz. Functional images were re-aligned using MCFLIRT \cite{jenkinson_improved_2002} and skull-stripped using BET \cite{smith_fast_2002}. Confound regression was performed using a 36-parameter model; confounds included global signal, 6 motion parameters as well as their temporal derivatives, quadratic terms, and the temporal derivatives of the quadratic terms \cite{satterthwaite_heterogeneous_2013}. Prior to confound regression, all confound parameters were band-pass filtered in a fashion identical to that applied to the original timeseries data, ensuring comparability of the signals in frequency content \cite{hallquist_nuisance_2013}. Subjects that displayed high levels of motion (mean relative RMS greater than 0.20 mm or more than 20 frames with over 0.25 mm of motion) or poor signal coverage were excluded from all analysis. Functional images were coregistered to the T1 image using boundary-based registration \cite{greve_accurate_2009} and aligned to template space using ANTs, as described above, all transforms were concatenated and thus only one interpolation was performed.

\subsubsection*{Network construction}

We extracted time-varing mean blood oxygen level-dependent signal from $N = 360$ regions of interest, which collectively comprised a multimodal parcellation of the cerebral cortex \cite{glasser_multi-modal_2016}. We estimated the functional connectivity \cite{friston2011functional} between any two brain regions $i$ and $j$ by calculating the Pearson correlation coefficient \cite{zalesky2012on} between the mean activity time series of region $i$ and the mean activity time series of region $j$ \cite{biswal_functional_1995}. We represented the $N \times N$ functional connectivity matrix as a graph or network in which regions were represented by network nodes, and in which the functional connectivity between region $i$ and region $j$ was represented by the network edge between node $i$ and node $j$. We used this encoding of the data as a network to produce an undirected, signed adjacency matrix $\mathbf{A}$. 

Original efforts in the emerging field of network neurosciece \cite{bassett2017network} began by studying binary graphs, where edges were assigned weights of either $1$ or $0$ \cite{sporns2005human,kaiser2006nonoptimal,scannell1995analysis}. In the context of functional brain networks, these early studies thresholded the functional connectivity matrices, often based on statistical testing of significance \cite{achard2006resilient}. Edges with original weights greater than the threshold were maintained as edges with a new weight of $1$, while edges with original weights less than the threshold were assigned a new weight of $0$ \cite{bassett2006adaptive,rubinov2009small,stam2007small}. However, recent evidence has demonstrated that the maintenance of edge weights is critical for an accurate understanding of the underlying biology of neural systems \cite{bassett2016small}. Simultaneously, recent evidence in applied mathematics has demonstrated that graph-related calculations are markedly more robust in weighted graphs than in binary graphs \cite{good2010performance}. In light of these two strands of evidence from the application domain and from mathematics, we maintained all edge weights without thresholding, and studied the full graph including both positive and negative correlations.

In the construction of these functional brain networks, we noticed that some participants did not have adequate coverage of all 360 brain regions. Because graph statistics tend to be heavily biased by the size of the graph, we ensured that the number of brain regions was consistent across participants. Specifically, we observed that more than 5 participants did not have adequate coverage of one region, leaving 359 nodes for subsequent analysis.

\subsection*{Quantification and Statistical Analysis} 
\subsubsection*{Network statistics}

In the course of our investigation, we sought to assess the effects of SES on both local and mesoscale architecture in functional brain networks estimated from fMRI BOLD measurements acquired at rest. To assess local network architecture, we used the most commonly studied graph measure of local connectivity -- the clustering coefficient -- which is commonly interpreted as reflecting the capacity of the system for processing within the immediate neighborhood of a given network node \cite{bartolomei2006disturbed,achard2006resilient,bassett2006adaptive,xu2016network}. While mesoscale network architecture comes in several forms \cite{betzel2018diversity}, we considered the most commonly studied mesocale organization -- assortative community structure -- which is commonly assessed by maximizing a modularity quality function \cite{porter2009communities,fortunato2010community}. Together, these two measures allow us to distinguish between effects of SES that are differentially located within immediate or extended neighborhoods of the functional brain network. 

\noindent \textbf{Clustering coefficient.} While the clustering coefficient has been defined in several different ways, it is generally considered to be a measure of local network segregation that quantifies the amount of connectivity in a node's immediate neighborhood. Intuitively, a node has a high clustering coefficient when a high proportion of its neighbors are also neighbors of each other. We specifically used a formulation that was recently generalized to signed weighted networks \cite{costantini_generalization_2014,zhang_general_2005}. This version is sensitive to non-redundancy in path information based on edge sign as well as edge weight, and importantly distinguishes between positive triangles and negative triangles, which have distinct meanings in networks constructed from correlation matrices.

To present the formal definition of the clustering coefficient, we begin by representing the functional connectivity network of a single participant as the graph $G = (V, E)$, where $V$ and $E$ are the vertex and edge sets, respectively. Let $a_{ij}$ be the weight associated with the edge $(i,j) \in V$, and define the weighted adjacency matrix of $G$ as $A = [a_{ij}]$. The clustering coefficient of node $i$ with neighbors $j$ and $q$ is given by
\begin{equation}
C_i =  \frac{\sum_{jq} (a_{ji} a_{iq} a_{jq}) }
{\sum_{j \neq q} | a_{ji} a_{iq} |}~~.
\end{equation}
\noindent The clustering coefficient of the entire network was calculated as the average of the clustering coefficient across all nodes as follows:
\begin{equation}
C = \frac{1}{n} \sum_{i \in N} C_i~~.
\end{equation}
\noindent In this way, we obtained estimates of the regional and global clustering coefficient for each subject in the sample.

\noindent \textbf{Modularity quality index.} As with the clustering coefficient, there are many different statistics that have been defined to quantify the modular structure of a network. Yet, many of them have in common the fact that they have been constructed to assess the extent to which a network's nodes can be subdivided into groups or \emph{modules} characterized by strong, dense intra-modular connectivity and weak, sparse inter-modular connectivity. Our approach is built on the modularity quality function originally defined Newman \cite{newman2006modularity}, and subsequently extended to weighted and signed networks by various groups. 

Specifically, we follow \cite{rubinov_weight-conserving_2011} by first letting the weight of a positive connection between nodes $i$ and $j$ be given by $a^+_{ij}$, the weight of a negative connection between nodes $i$ and $j$ be given by $a^-_{ij}$, and the strength of a node $i$, $s^\pm_i = \sum_j a^{\pm_{ij}}$, be given by the sum of the positive or negative connection weights of $i$. We denote the chance expected within-module connection weights as $e^+_{ij}$ for positive weights and $e^-_{ij}$ for negative weights, where $e^\pm_{ij} = \frac {s^\pm_i s^\pm_j}{v^\pm}$. We let the total weight, $v^\pm = \sum_{ij} a^\pm_{ij}$, be the sum of all positive or negative connection weights in the network. Then the asymmetric generalization of the modularity quality index is given by:
\begin{equation}
Q^* = \frac{1}{v^+} \sum_{ij}(a^+_{ij} - e^+_{ij}) \delta_{M_i M_j} - \frac{1}{v^+ + v^-} \sum_{ij}(a^-_{ij} - e^-_{ij}) \delta_{M_i M_j}~,
\end{equation}
where $M_{i}$ is the community to which node $i$ is assigned, and $M_{j}$ is the community to which node $j$ is assigned. We use a Louvain-like locally greedy algorithm as a heuristic to maximize this modularity quality index subject to a partition $M$ of nodes into communities.

\subsubsection*{Network null models}

To explicitly test whether the topological properties of functional brain networks were significantly different from those that would be expected by chance, we conducted comparisons with two distinct random network null models that differed in their level of stringency. To define these null models, we first note that a node's degree is given by the number of edges emanating from it or leading to it, and the strength of a node is the average weight of the edges emanating from it or leading to it. Our first, and least stringent null model, was one in which the empirical network topology is destroyed by permuting the location of edges uniformly at random while maintaining the degree distribution. This null model is commonly used in the literature, but may be better suited to the study of binary networks where the degree distribution is an important feature, than it is to the study of weighted networks where the strength distribution may also be an important feature of the topology \cite{bassett2016small,rubinov_weight-conserving_2011}. To be conservative therefore, we also employed a null model that preserved both the degree and strength distributions. We generated a total of 100 instantiations of each null model per participant. 

\subsubsection*{Regional homogeneity}

To gain a fuller understanding of the neural mechanisms underlying developmental changes in functional network topology, we also considered the local organization of the BOLD signal within each region of interest. Specifically, we estimated voxelwise regional homogeneity (ReHo) using Kendall's coefficient of concordance computed over the BOLD timeseries in each voxel's local neighborhood, defined to include the 26 voxels adjoining its faces, edges, and vertices \cite{zang_regional_2004}. Preprocessing steps were identical to those described above. We then computed the mean ReHo of each of the $N=360$ cortical regions of interest. 

\subsubsection*{Statistical modeling and testing}

The relationship between various markers of brain development and age is sometimes nonlinear \cite{shaw_intellectual_2006}, as is the relationship between these same markers and SES \cite{piccolo_age-related_2016}. Thus, we first examined our data for the presence of nonlinear relationships between the clustering coefficient and age. Flexible nonlinear functions were estimated using generalized additive models (GAMs) with the \emph{mgcv} package in R \cite{satterthwaite_impact_2014, wood_fast_2011}. The penalty parameters for the nonlinear spline terms were fit as random effects and tested using restricted likelihood ratio tests with RLRsim \cite{size_scheipl_2008}. Note that these tests of nonlinearity are constructed so as to test for nonlinear effects over and above any linear effects that may be present. 

After confirming the absence of any significant nonlinear relationships, we proceeded with linear models. Specifically, we modeled the linear effect of age and the age $\times$ SES interaction on network topology while controlling for sex, race, head motion (mean relative RMS displacement over the whole timeseries), and average edge weight in the functional brain network. The choice to include the average edge weight as a covariate of noninterest ensures that subsequent results reflect changes in network topology rather than global differences in connectivity strength \cite{yan_standardizing_2013, wijk_comparing_2010, ginestet_brain_2011}. We used multiple ordinary least squares (OLS) linear regression with the \emph{lm()} command in R to fit the following general equation:
\begin{equation}
C = \mathrm{age} + \mathrm{sex} + \mathrm{race} + \mathrm{mean~RMS} + \mathrm{weight} + \mathrm{SES} + \mathrm{age} * \mathrm{SES}~,
\end{equation}
where $C$ is the global clustering coefficient. Analyses examining the main effect of age or SES on network topology omitted the interaction term. All $\beta$ values reported are standardized coefficients. We used the R package \emph{visreg} to calculate 95\% confidence intervals around fitted lines and to generate partial residuals.

We used an identical model when estimating age and age $\times$ SES effects on network statistics of individual brain regions (Fig. 4). For all analyses performed at the node level, we controlled for multiple comparisons using the False Discovery Rate ($q < 0.05$) \cite{benjamini_controlling_1995}. We tested for significant differences between nested models when appropriate using likelihood ratio tests.

To examine effects of SES on development within putative functional networks, we assigned each node to one of seven large-scale functional systems defined \emph{a priori} \cite{yeo_organization_2011}. We then calculated the average clustering coefficient across nodes within each system and examined the linear effect of age and the age $\times$ SES interaction for each system separately. To examine whether the age $\times$ SES interaction differed across cognitive systems, we estimated the similarity between the system partition and the distribution of the effect estimates. Specifically, we first normalized and discretized the effect estimates, and then we calculated the $z$-score of the Rand coefficient between the discretized interaction effects across nodes and the partition of nodes into systems \cite{traud_comparing_2011}. We assessed statistical significance using a nonparametric permutation test; we permuted the distribution of interaction estimates uniformly at random 10,000 times, and for each permutation calculated the $z$-score of the Rand coefficient between the interaction estimates and the Yeo system partition. Then, we rank-ordered the $z$-scores of the Rand coefficient on permuted data and compared the true value of the $z$-score of the Rand coefficient to that of the null distribution.

To examine the relationship between regional homogeneity and the clustering coefficient, we examined pairwise partial correlations between the two metrics across regions, controlling for all covariates specified in \textbf{Eq.~4} above. To examine age and age $\times$ SES effects on the clustering coefficient above and beyond the effects of ReHo, we regressed the regional ReHo out of the regional clustering coefficient, and used the residuals of this regression as the variable of interest in \textbf{Eq.~4} in place of the raw clustering coefficient to again assess effects across cortical areas.

To examine the effect of physical distance on our findings, we follow \cite{betzel2018specificity} by thresholding each network by distance in bins, ranging from the top 10\% of shortest edges to the bottom 10\% of longest edges. We then calculated the clustering coefficient on these sparse networks, and estimated the age $\times$ SES effect within each distance bin. The physical distance of an edge was estimated as the Euclidean distance between the centroids of two regions of interest in the template space. Although Euclidean distance is an imperfect proxy for anatomical connection distance, previous work has shown it to be comparable to measures of wiring distance based on diffusion imaging data \cite{supekar_development_2009}. We assessed statistical significance using null network models that permuted the distribution of edge weights uniformly at random within subgraphs. We calculated the clustering coefficient on these null network models, and we compared the age $\times$ SES effect observed in the null models to that observed in the empirical data.

We used non-parametric permutation tests to assess the significance of the differences in strength of age $\times$ SES effects across edges. We permuted the distribution of edge weights uniformly at random and calculated standardized coefficients for the age $\times$ SES effect, averaged for (i) edges between nodes that demonstrated a significant node-level effect, (ii) edges between these nodes and the rest of the brain, and (iii) edges connecting nodes that did not show a significant node-level effect. We then compared the true differences in standardized coefficients to the differences calculated from permuted distributions of edge weights.

We conducted all analyses in R \cite{R_core_team_2013} and MATLAB using custom code as well as functions from the Brain Connectivity Toolbox \cite{rubinov_complex_2010}.

\subsection*{Data and Software Availability}
The data reported in this paper have been deposited in the database of Genotypes and Phenotypes under accession number dbGaP: phs000607.v2.p2 (\url{https://www.ncbi.nlm.nih.gov/projects/gap/cgi-bin/study.cgi?study_id=phs000607.v2.p2}). Publicly available MATLAB code used to calculate network measures is available in the Brain Connectivity Toolbox \cite{rubinov_complex_2010}. Code for analyses presented here is available at \url{www.github.com/utooley/rsfmri_envi_networks}.

\end{document}


\title{Supplementary Materials and Methods for\\ ``Influence of Neighborhood SES on Functional Brain Network Development''}
\maketitle
\date{\today}

\section{Neighborhood SES modeled continuously}
In the main text we conducted a median split on neighborhood SES for ease of presentation and interpretation. Here we conduct parallel analyses with SES modeled as a continuous variable, and find that the results are qualitatively unchanged for the measure of local segregation (clustering coefficient). We observed a significant interaction between SES as a continuous variable and age, such that higher SES youth demonstrated a more rapid increase in average clustering coefficient with age than lower SES youth ($p=0.0489$). Age ($\beta = 0.17$, $p < 1 \times 10^{-7}$), mean edge weight ($\beta = 0.17$, $p < 1 \times 10^{-6}$), sex ($\beta=0.10$,$p= 0.001$), and motion ($\beta = -0.17$, $p < 1 \times 10^{-7}$) were also significant predictors of the clustering coefficient in this model (full model: $F(8, 1003) = 14.68$, $R^{2} = 0.10$, $p < 1 \times 10^{-15}$). 

We did not observe a significant interaction between neighborhood SES modeled continuously and age for the modularity quality index, our meso-scale metric of segregation ($p > 0.12$).

\section{Effects of Maternal Education on Local Segregation}
A total of $n=999$ participants had information on maternal education. Within this sample, we observed that maternal education was highly correlated with neighborhood SES (Spearman's $\rho=0.48$, $p < 1 \times 10^{-15}$). Therefore, we included maternal education in our multiple regression model to determine any potential effects on our results and conclusions. We observed that our results were qualitatively unchanged. Age remained a significant predictor of the average clustering coefficient ($\beta = 0.17$, $p < 1 \times 10^{-7}$), with sex ($\beta = 0.09$, $p = 0.002$), mean edge weight ($\beta = 0.17$, $p < 1 \times 10^{-6}$), and in-scanner motion ($\beta = -0.18$, $p < 1 \times 10^{-7}$) also remaining significant predictors (full model: $F(8, 990) = 13.83$, $R^{2} = 0.09$, $p < 1 \times 10^{-15}$). A model including the interaction between neighborhood SES and age was also significant ($p = 0.005$, full model: $F(9, 989) = 13.26$, $R^{2} = 0.10$, $p < 1 \times 10^{-15}$). The results of these analyses indicate that neighborhood SES may be capturing aspects of variation in the environment that are uniquely related to developmental increases in local segregation in youth.

We also examined our data for the presence of an age-by-maternal-education interaction. We found no significant interaction between age and maternal education in predicting either local-scale or meso-scale segregation ($p$-values $> 0.41$). 

\section{Sensitivity Analysis}
\begin{figure}
	\includegraphics[width=0.6\textwidth]{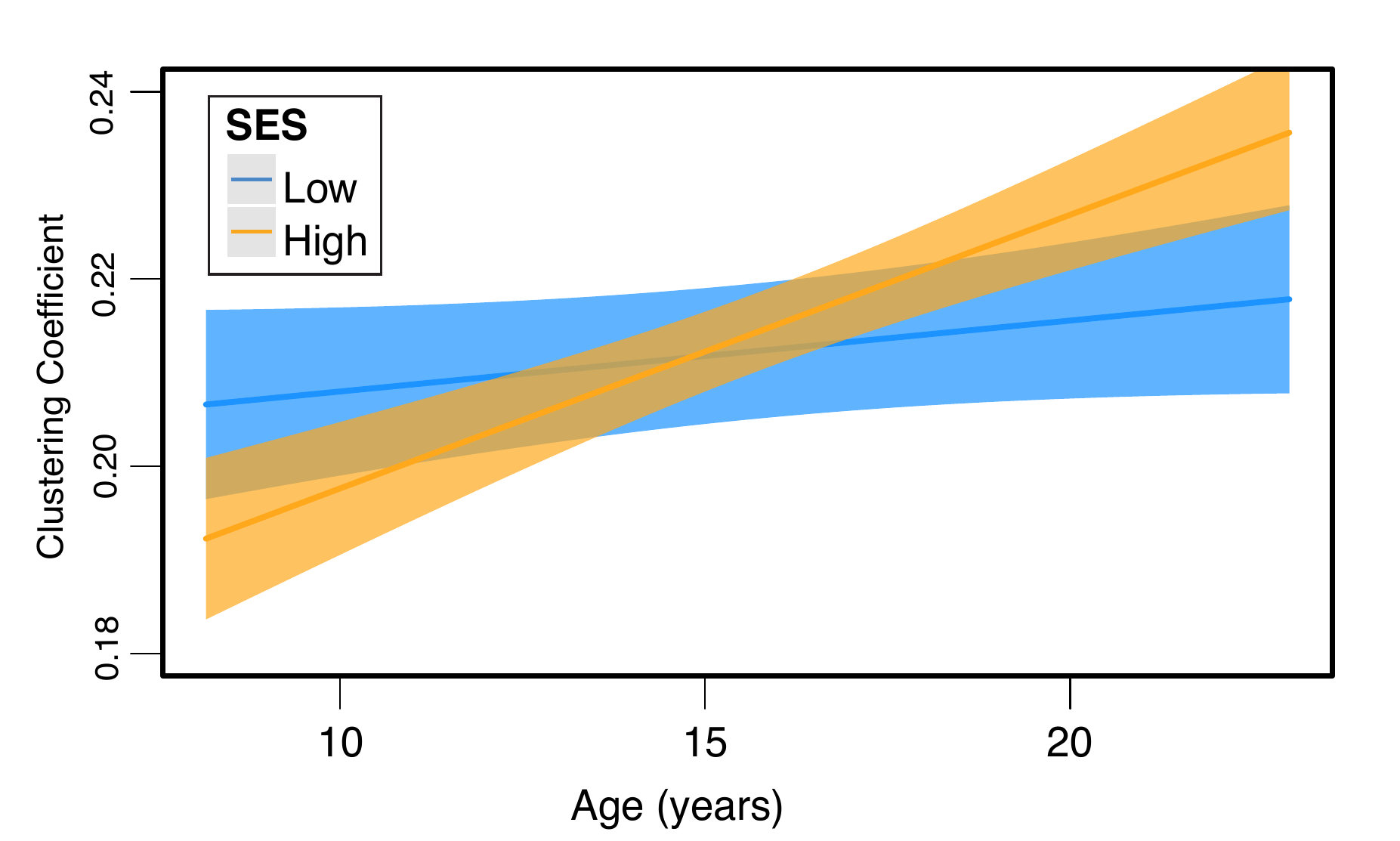}
\caption{\textbf{Sensitivity analysis conducted with a smaller sample}. We conducted sensitivity analyses with a smaller sample ($n=883$), after excluding those participants that were currently taking psychoactive medication or that had a history of psychiatric hospitalization. Results were qualitatively similar to those found in the full sample (see main text), with high-SES youth displaying a swifter increase in average clustering coefficient over this developmental period than low-SES youth.}
\label{supfig3}
\end{figure}

We conducted sensitivity analyses with a smaller sample ($n=883$), after excluding participants that were currently taking psychoactive medication or who had a history of psychiatric hospitalization. Results were qualitatively similar to those found in the full sample (see \textbf{Sup. Fig. ~\ref{supfig3}}). We observed no significant nonlinear relationship between the whole-brain clustering coefficient and age ($\mathrm{RLRT} = 0.60$, $p > 0.13$), and we found that age significantly predicted whole-brain clustering coefficient ($\beta = 0.17$, $p < 1 \times 10^{-6}$). We also observed that sex ($\beta = 0.08$, $p = 0.02$), mean edge weight ($\beta = 0.17$, $p < 1 \times 10^{-6}$), and in-scanner motion ($\beta = -0.16$, $p < 1 \times 10^{-5}$) were significant predictors of the average clustering coefficient. The main effect of SES was not significant ($\beta = 0.04$, $p =0.4$, full model: $F(7, 875) = 12.43$, $R^{2} = 0.08$, $p < 1 \times 10^{-14}$). 

We observed a significant interaction between SES and age, such that high-SES youth displayed a swifter increase in mean clustering coefficient with age than low-SES youth ($p = 0.004$). We also observed that sex ($\beta = 0.08$, $p = 0.01$), mean edge weight ($\beta = 0.16$, $p < 1 \times 10^{-5}$) and motion ($\beta = -0.15$, $p < 1 \times 10^{-5}$) were significant predictors of the clustering coefficient in this model (full model: $F(8, 874) = 12.01$, $R^{2} = 0.09$, $p < 1 \times 10^{-16}$). 

\section{Replication in an Alternative Parcellation}
To further validate our results, we employed an alternative parcellation generated from gradients and similarity of intrinsic functional connectivity patterns \cite{schaefer_local-global_2017}. Specifically, we extracted time-varing mean blood oxygen level dependent signal from $N = 400$ regions of interest in this alternative parcellation, and repeated our primary analyses on the effect of age and SES on whole-brain functional topology at rest ($n=1015$, all subjects had adequate coverage of regions in this parcellation). As in our previous analyses, we modeled the linear effect of age and the age $\times$ SES interaction on network topology while controlling for sex, race, head motion (mean relative RMS displacement over the whole timeseries), and average edge weight in the functional brain network (see \textbf{Eq.~4} in the main text). 

\begin{figure}
	\includegraphics[width=\textwidth]{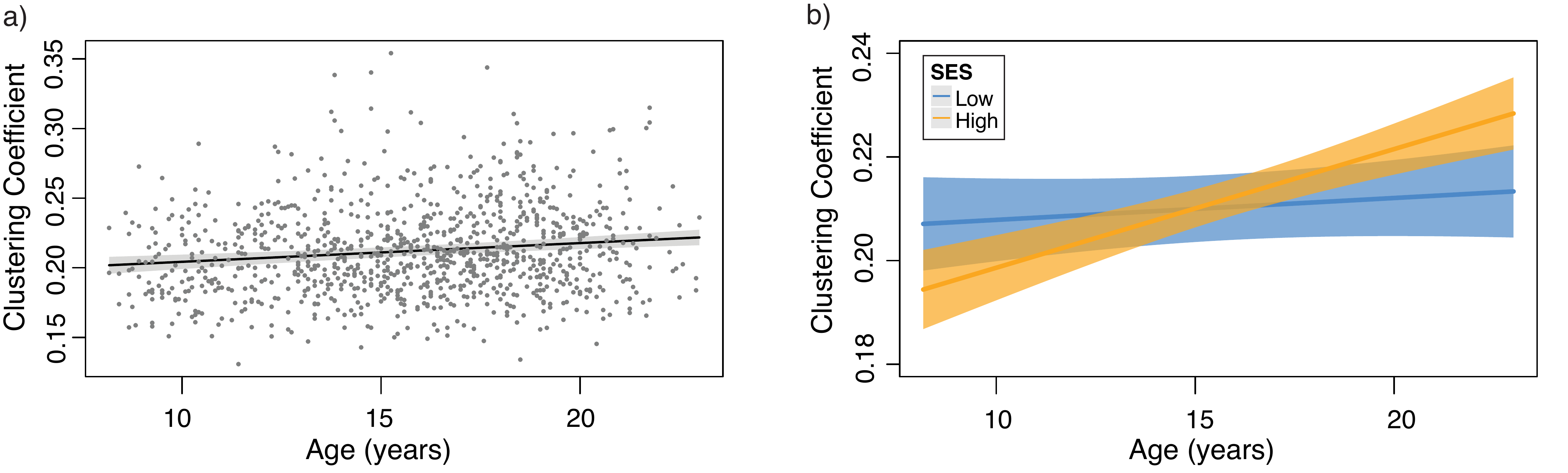}
\caption{\textbf{Replication in an alternative parcellation.} \emph{(a)} Average clustering coefficient increases with age in an alternative parcellation, after controlling for sex, race, head motion, and mean edge weight. Note that values plotted are partial residuals. \emph{(b)} High-SES youth display a swifter increase in the average clustering coefficient over this developmental period than low-SES youth.}
\label{supfig4}
\end{figure}

Results were qualitatively similar to those reported in the main text. We found that age significantly predicted average clustering coefficient: older children had higher average clustering coefficients than younger children (\textbf{Sup. Fig.~\ref{supfig4}a}, $\beta = 0.13$, $p < 1 \times 10^{-4}$). Furthermore, sex ($\beta = 0.07$, $p = 0.01$), mean edge weight ($\beta = 0.28$, $p < 1 \times 10^{-15}$), and in-scanner motion ($\beta = -0.20$, $p < 1 \times 10^{-10}$) were also significant predictors of average clustering. The main effect of SES was not significant ($\beta = 0.02$, $p =0.66$; full model: $F(7, 1007) = 25.78$, $R^{2} = 0.15$, $p < 1 \times 10^{-15}$). We observed no significant non-linear relationship between the average clustering coefficient and age (restricted likelihood ratio test $\mathrm{RLRT} = 0.29$, $p > 0.17$) or average network weight ($p > 0.8$). We also observed a significant interaction between SES and age, such that high-SES youth demonstrate a more rapid increase in average clustering coefficient with age than low-SES youth (\textbf{Sup. Fig.~\ref{supfig4}b}, $p =0.002$). Mean edge weight ($\beta = 0.16$, $p < 1 \times 10^{-15}$), sex ($\beta = 0.16$, $p = 0.01$), and motion ($\beta = -0.17$, $p < 1 \times 10^{-10}$) were also significant predictors of the clustering coefficient in this model (full model: $F(8, 1006) = 23.95$, $R^{2} = 0.15$, $p < 1 \times 10^{-15}$). 

Additionally, effects on the modularity quality index were qualitatively similar to those reported in the main text. Specifically, we observed that the modularity quality index increased with age ($\beta = 0.14$, $p < 1 \times 10^{-4}$). Furthermore, we observed that sex ($\beta = 0.08$, $p = 0.01$), mean edge weight ($\beta = -0.16$, $p < 1 \times 10^{-6}$), and in-scanner motion ($\beta = -0.14$, $p < 1 \times 10^{-4}$) were all significant predictors of the modularity quality index. The main effect of SES was not significant ($\beta = -0.03$, $p =0.51$; full model: $F(7, 1007) = 12.23$, $R^{2} = 0.07$, $p < 1 \times 10^{-14}$). A model including the interaction of age and SES was also significant ($F(8, 1006) = 11.70$, $R^{2} = 0.08$, $p < 1 \times 10^{-15}$). The age-by-SES interaction was significant ($p = 0.006$), and the mean edge weight ($\beta = -0.17$, $p < 1 \times 10^{-7}$), sex ($\beta = 0.08$, $p = 0.01$), and motion ($\beta = -0.13$, $p < 1 \times 10^{-4}$) were also significant predictors of the modularity quality index in this model.

\section{Regional Homogeneity Models}

ReHo measurements averaged over the whole brain showed a main effect of age: ReHo significantly decreased with age ($\beta = -0.14$, $p < 1 \times 10^{-5}$). Interestingly, race ($\beta = 0.16$, $p = 0.001$) and sex ($\beta = -0.08$, $p =0.01$) were also significant predictors of ReHo. The main effect of SES was not significant ($\beta = -0.03$, $p = 0.5$). In a model including the age-by-SES effect, the interaction was not significant ($\beta = 0.04$, $p = 0.33$). 

\newpage
\begin{figure*}
	\includegraphics[width=0.49\linewidth]{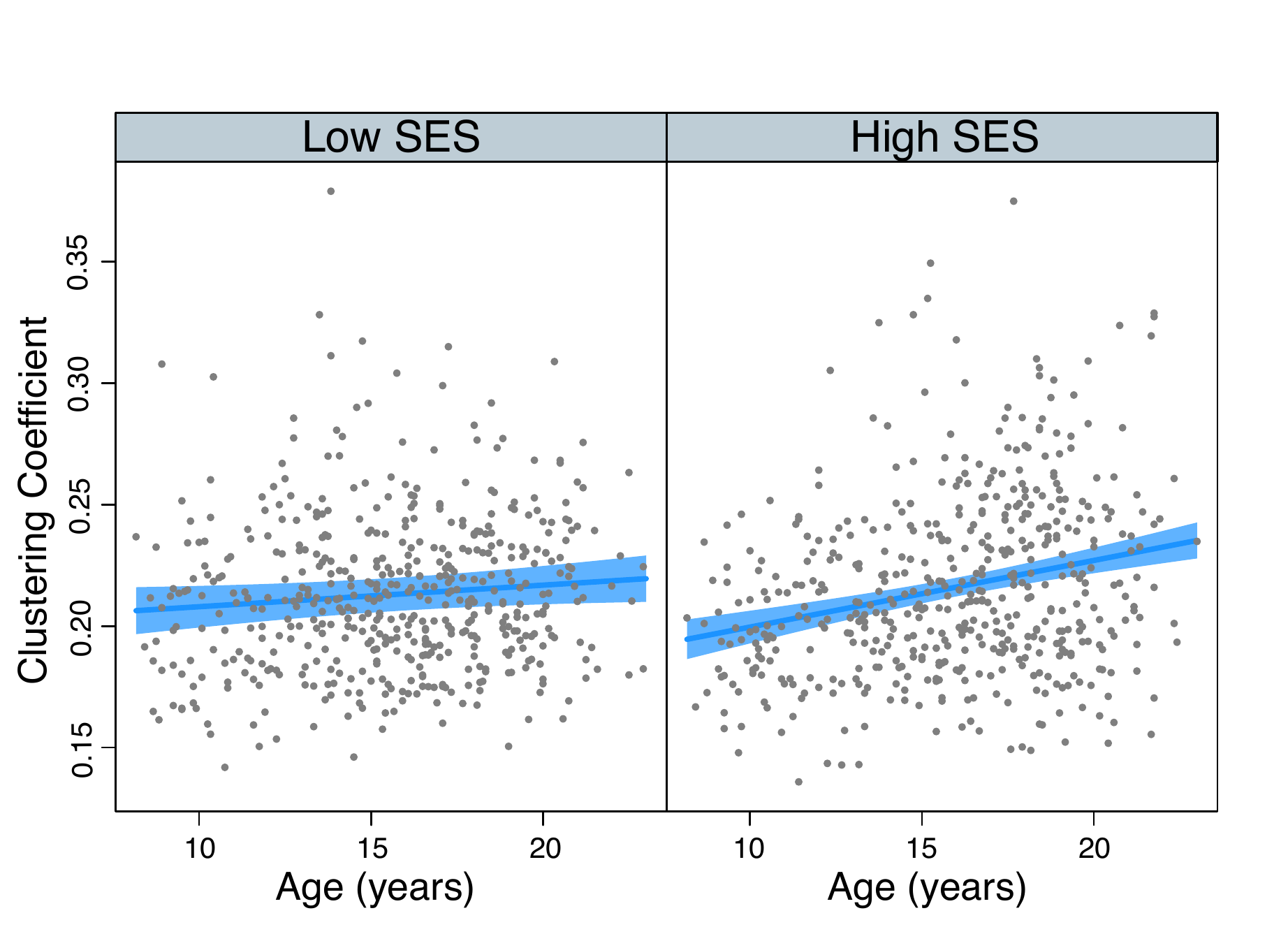} 
\caption{\textbf{Effects of age and SES on whole-brain average clustering coefficient}. High-SES youth display a swifter increase in average clustering coefficient than low-SES youth over this developmental period. Figure displays all datapoints, and values plotted are partial residuals.}
\label{supfig1}
\end{figure*}

\begin{figure*}
	\includegraphics[width=\linewidth]{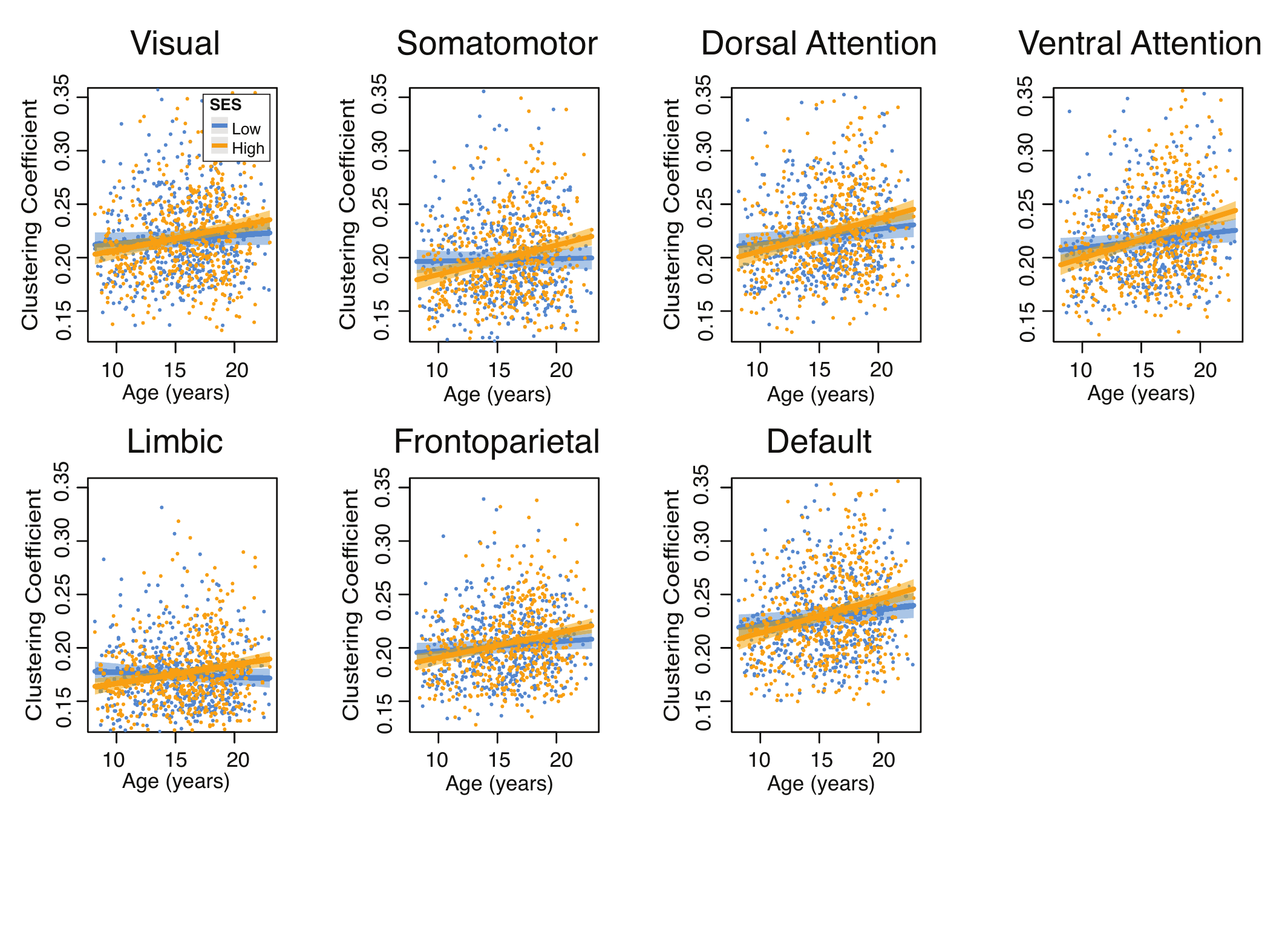} 
\caption{\textbf{Effects of age and SES on systems-level functional network topology}. Figure displays age $\times$ neighborhood SES interaction for each cognitive system separately. Effects of age on increases in the clustering coefficient are strongest in the default mode, ventral attention, and dorsal attention systems. SES effects on age-related increases in the clustering coefficient were strongest in the limbic, somatomotor, and ventral attention systems. Some systems that demonstrate weaker age effects show stronger interaction effects, perhaps due to the limited range of developmental change in low-SES youth. Figures display all datapoints, and values plotted are partial residuals.}
\label{supfig2}
\end{figure*}

\begin{figure}
	\includegraphics[width=0.5\textwidth]{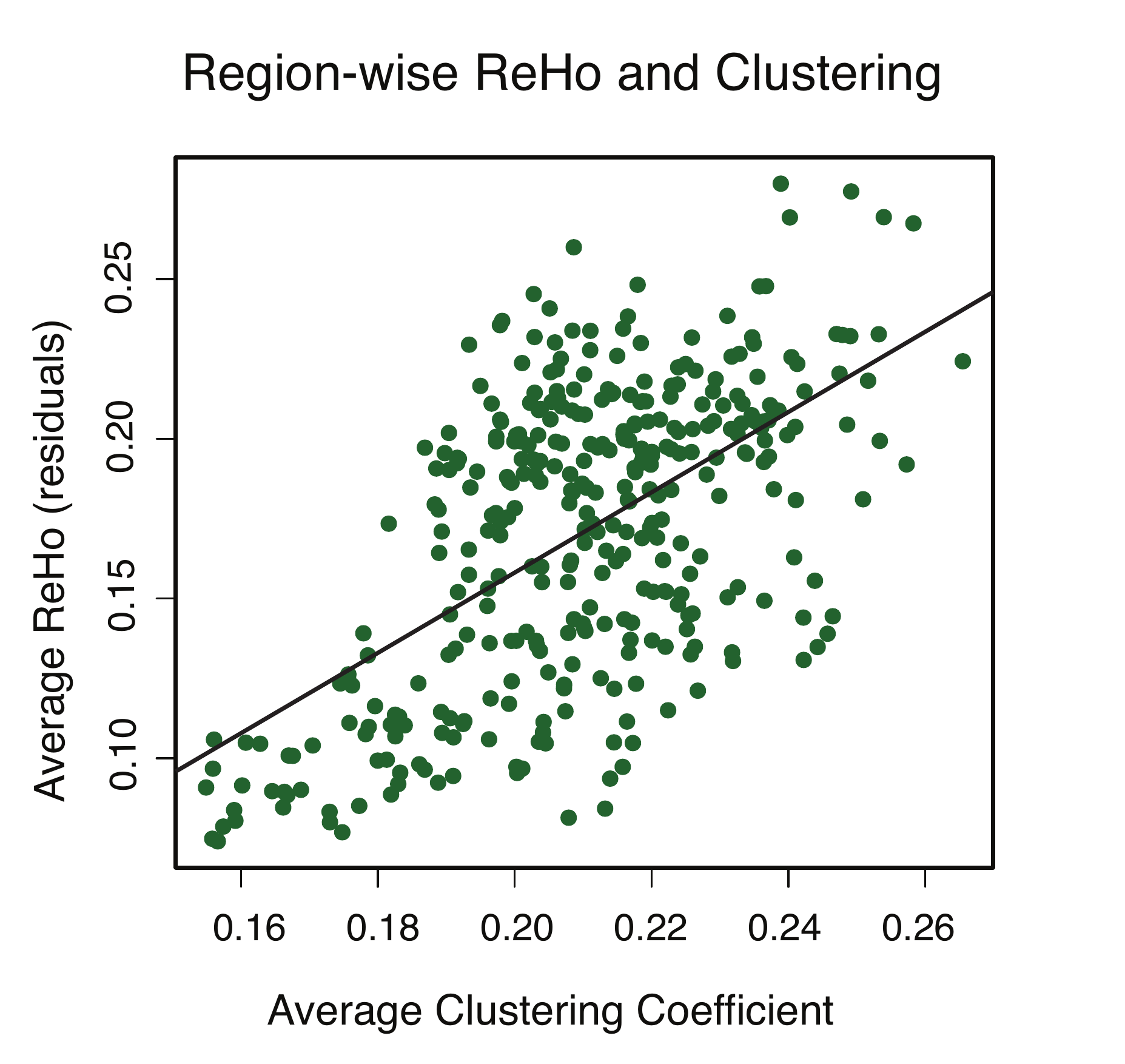}
\caption{\textbf{Relationship between region clustering coefficient and regional homogeneity.} Across regions, ReHo values are highly correlated with the clustering coefficient when controlling for age, sex, motion, and average network weight. Note that values plotted are partial residuals.}
\label{supfig5}
\end{figure}

\clearpage

\begin{table}[]
\centering
\caption{\textbf{Bivariate Relationships Among Predictors of Interest}}
\label{sup_table_1}
\begin{tabular}{rrrrrrr}
               & Age (years) & Network weight & Mean RMS & SES   & Clustering & Q      \\
Age (years)    &             & $0.08^{*}$          & $-0.17^{***}$   & 0.03  & $0.21^{***}$  & $0.15^{***}$  \\
Network weight &             &                & $0.22^{***}$    & -0.01 & $0.12^{***}$      & $-0.18^{***}$ \\
Mean RMS       &             &                &          & 0.00  & $-0.19^{***}$     & $-0.15^{***}$ \\
SES            &             &                &          &       & $0.06$      & $0.05$  \\
Clustering     &             &                &          &       &            & $0.78^{***}$  \\
Q              &             &                &          &       &            &      \\
\hline 
\hline \\[-1.8ex] 
\end{tabular}
\caption*{\textit{Note: Spearman's $\rho$ for pairwise correlations between variables. Clustering: Average whole brain clustering coefficient. Q: Modularity quality index. $^{*}$$p < 0.05$; $^{**}$$p < 0.01$; $^{***}$$p < 0.001$}}
\end{table}

\begin{table}[!htbp] \centering 
  \caption{\textbf{Full model for whole brain clustering coefficient.}} 
  \label{sup_table_2} 
\begin{tabular}{@{\extracolsep{4pt}}lrrrcrrrc}
\\[-1.8ex]\hline 
\hline \\[-1.8ex] 
 & \multicolumn{8}{c}{\textit{Whole Brain Clustering Coefficient}} \\ 
\\[-1.8ex] & B(SE) & $\beta$ & $t$ & $p$ & B(SE) & $\beta$ & $t$ & $p$ \\ 
\hline \\[-1.8ex] 
 Age (years) & 0.002 (0.0003) & 0.17 & 5.438 & $<$ 0.001$^{***}$  & 0.002 (0.0003) & 0.17 & 5.472 & $<$ 0.001$^{***}$ \\ 
  Sex & 0.007 (0.002) & 0.10 & 3.202  &  0.002$^{**}$  & 0.007 (0.002) & 0.10 & 3.183 &  0.002$^{**}$ \\ 
  Race (Black) & $-$0.004 (0.003) & -0.05 & $-$1.060 &  0.289 & $-$0.003 (0.003) & -0.05 & $-$0.978 &  0.328\\ 
  Race (Other) & 0.003 (0.004) & 0.03 & 0.785 &  0.433 & 0.003 (0.004) & 0.03 & 0.853 &  0.394\\ 
  Network weight & 0.837 (0.155)  & 0.17 & 5.419  & $<$ 0.001$^{***}$  & 0.812 (0.154) & 0.17 & 5.261& $<$ 0.001$^{***}$ \\ 
  Mean RMS & $-$0.173 (0.031) & -0.18 & $-$5.630 & $<$ 0.001$^{***}$  & $-$0.168 (0.031) & -0.17 & $-$5.472 & $<$ 0.001$^{***}$  \\ 
  SES & 0.001 (0.002) & 0.03 & 0.649 &  0.517 & $-$0.019 (0.008) & -0.37 & $-$2.523 &  0.012$^{*}$ \\ 
  Age x SES & & & & & 0.001 (0.0005) & 0.41 & 2.857 &  0.004$^{**}$ \\ 
  Intercept & 0.174 (0.007) & & 26.144 & $<$ 0.001$^{***}$  & 0.174 (0.007) & & 26.212 & $<$ 0.001$^{***}$ \\
 \hline \\[-1.8ex] 
R$^{2}$ & & & & 0.101 & & & & 0.108 \\ 
Adjusted R$^{2}$ & & & & 0.094 & & & & 0.101 \\ 
F Statistic & & & 16.070$^{***}$ &(df = 7; 1004) & & & 15.182$^{***}$ & (df = 8; 1003) \\ 
\hline 
\hline \\[-1.8ex] 
& \multicolumn{8}{r}{\textit{Note:$^{*}$$p < 0.05$; $^{**}$$p < 0.01$; $^{***}$$p < 0.001$}} \\ 
\end{tabular} 

\end{table} 

\bibliography{supplement_v1Notes}
\bibliographystyle{unsrt}